\documentclass[a4paper,11pt,nohyper]{JHEP3}
\usepackage{amsthm}
\newtheorem*{theorem}{Theorem}
\newtheorem*{lemma}{Lemma}

\def\a{\alpha}
\def\b{\beta}

\def\d{\delta}
\def\e{\epsilon}           

\def\t{\tau}

\def\z{\zeta}

\def\S{\Sigma}

\def\del{\partial}              
\let\a=\alpha \let\b=\beta  \let\d=\delta \let\e=\epsilon
\let\z=\zeta    
    
 \let\t=\tau

\def\nn{\nonumber} \def\bd{\begin{document}} \def\ed{\end{document}}
\def\ds{\documentstyle} \let\fr=\frac \let\bl=\bigl \let\br=\bigr
\let\Br=\Bigr \let\Bl=\Bigl
\let\bm=\bibitem
\let\na=\nabla
\let\pa=\partial \let\ov=\overline
\newcommand{\be}{\begin{equation}}
\newcommand{\ee}{\end{equation}}
\def\ba{\begin{array}}
\def\ea{\end{array}}
\def\ft#1#2{{\textstyle{{\scriptstyle #1}\over {\scriptstyle #2}}}}
\def\fft#1#2{{#1 \over #2}}
\def\del{\partial}
\def\sst#1{{\scriptscriptstyle #1}}
 \def\oneone{\rlap 1\mkern4mu{\rm l}}
\def\ie{{\it i.e.\ }}
\def\via{{\it via}}
\def\semi{{\ltimes}}
\def\str{{\rm str}}
\def\Dm{{{D_{\sst{max}}}}}
\def\vac{ \left | 0 \right \rangle }
\def\kvac{ \left | k \right \rangle }

\def\sp{\; \; \;}

\def\bol{ \left | B (p^+) \right \rangle}
\def\bo1{ \left | B^0 (p^+) \right \rangle}

\def\bolt{ \left | B (p^+) \right \rangle_{\t}}

\def\boxl{ \left | B (x^-) \right \rangle}
\newcommand{\bea}{\begin{eqnarray}}
\newcommand{\eea}{\end{eqnarray}}

\def\<{ \langle }
\def\>{ \rangle }

\def\S{\Sigma}

\usepackage{amsmath,latexsym,amssymb,slashed}
\usepackage{graphicx}
\usepackage[latin1]{inputenc}
\usepackage{subfigure}

\renewcommand{\floatpagefraction}{0.6}
\renewcommand{\textfraction}{0.2}

\newcommand\ca{\mathcal{A}}
\newcommand\vp{\varphi}
\newcommand\beal{\begin{align}}
\newcommand\bbone{\ensuremath{\mathbbm{1}}}

\newcommand{\eq}[1]{\begin{equation}#1\end{equation}}
\newcommand{\spl}[1]{\begin{split}#1\end{split}}
\newcommand{\al}[1]{\begin{align}#1\end{align}}
\newcommand{\subeq}[1]{\begin{subequations}#1\end{subequations}}

\newcommand{\arXividhepth}[1]{\href{http://arxiv.org/abs/#1}arXiv:{\tt #1} [hep-th]}
\newcommand{\arXividother}[2]{\href{http://arxiv.org/abs/#1}arXiv:{\tt #1} [#2]}

\newcommand{\bg}[1]{\hat{#1}}
\newcommand{\wj}{\widetilde{J}}
\newcommand{\reo}{\mathrm{Re}~\!\omega}
\newcommand{\imo}{\mathrm{Im}~\!\omega}
\newcommand{\ads}{AdS_4}
\newcommand{\mcal}{\mathcal{M}}
\newcommand{\ccal}{\mathcal{C}}
\newcommand{\ncal}{\mathcal{N}}
\newcommand{\boxedeq}[1]{
\begin{equation}
\fbox{
\rule[0.7cm]{0pt}{0pt}
$#1$
\rule[-0.45cm]{0pt}{0pt}
}
\end{equation}
}

\def\d{\text{d}}

\def\slashchar#1{\setbox0=\hbox{$#1$}           
\dimen0=\wd0                                 
\setbox1=\hbox{/} \dimen1=\wd1               
\ifdim\dimen0>\dimen1                        
\rlap{\hbox to \dimen0{\hfil/\hfil}}      
#1                                        
\else                                        
\rlap{\hbox to \dimen1{\hfil$#1$\hfil}}   
/                                         
\fi}

\def\Re           {{\rm Re\hskip0.1em}}
\def\Im           {{\rm Im\hskip0.1em}}

\newcommand{\E}{\text{\tiny E}}

\newcommand{\tV}{{\widetilde{V}}}
\newcommand{\tH}{{\tilde{h}}}
\newcommand{\tm}{{{m}}}
\newcommand{\tmu}{{\tilde{\mu}}}
\newcommand{\trho}{{\tilde{\rho}}}
\newcommand{\tv}{{\tilde{v}}}
\newcommand{\calo}{\mbox{${\cal O}$}}
\newcommand{\cala}{\mbox{${\cal A}$}}
\newcommand{\dd}{\mathrm{d}}
\newcommand{\ra}{\rightarrow}
\newcommand{\calv}{\mbox{${\cal V}$}}
\newcommand{\calh}{\mbox{${\cal H}$}}
\newcommand{\calm}{\mbox{${\cal M}$}}
\newcommand{\abs}[1]{\left| #1 \right|}
\newcommand{\zetaa}{{\psi}}
\newcommand{\tr}{{\rm tr}\,}

\newcommand{\ky}[1]{{\color{blue}{#1}}}

\title{Inhomogeneity simplified}

\author{Marika Taylor and William Woodhead
\\

School of Mathematical Sciences and STAG, \\
University of Southampton, \\
Highfield, Southampton, SO17 1BJ, UK.

\bigskip
 E-mail:
 \email{m.m.taylor@soton.ac.uk; w.woodhead@soton.ac.uk}}

\abstract{
We study models of translational symmetry breaking in which inhomogeneous matter field profiles can be engineered in such a way that black brane metrics remain isotropic and homogeneous. We explore novel Lagrangians involving square root terms and show how these are related to massive gravity models and to tensionless limits of branes. Analytic expressions for the DC conductivity and for the low frequency scaling of the optical conductivity are derived in phenomenological models, and  the optical conductivity is studied in detail numerically. 
The square root Lagrangians are associated with linear growth in the DC resistivity with temperature and also lead to minima in the optical conductivity at finite frequency, suggesting that our models may capture many features of heavy fermion systems.

}

\begin{document}

\newcommand{\td}{\tilde}
 \newcommand{\bc}{\begin{center}}
 \newcommand{\ec}{\end{center}}
 \newcommand{\bfr}{\begin{flushright}}
 \newcommand{\efr}{\end{flushright}}
 \newcommand{\bfl}{\begin{flushleft}}
 \newcommand{\efl}{\end{flushleft}}
 \newcommand{\bt}{\begin{tabular}}
 \newcommand{\et}{\end{tabular}}
 
\pagebreak 

\section{Introduction}

Holographic modelling of strongly coupled condensed matter systems has generated a great deal of interest over recent years; for reviews see \cite{Hartnoll:2011fn,Iqbal:2011ae}. 
It is remarkable that many features of strongly coupled matter can be captured by static, isotropic solutions of Einstein-Maxwell-dilaton models. Nonetheless as one tries to develop more realistic models it is clear that such holographic geometries cannot adequately capture many important features of strongly interacting systems.

The focus of this paper will be on modelling systems with broken spatial translational symmetry. Realistic condensed matter systems never have perfect translational symmetry: the symmetry is explicitly broken both by lattice effects and by the presence of inhomogeneities. This breaking of translational invariance is necessary for particles to dissipate momentum, without which there would be a delta function in the conductivity at zero frequency. 

 Diffeomorphism invariance of a field theory implies conservation of the stress energy tensor $T_{ij}$ via the diffeomorphism Ward identity. If one considers a field theory which has a conserved current $J_i$ and a scalar operator ${\cal O}$ then diffeomorphism invariance is violated whenever there is a position dependent source $A_i$ for the current $J_i$ or a similar source $\phi$ for the scalar operator, and the corresponding operators acquire expectation values. The diffeomorphism Ward identity takes the form
\be
\nabla ^i \langle T_{ij} \rangle -\langle J^i \rangle F_{ij} + \langle {\cal O} \rangle \partial_j \phi = 0,
\ee
with $F_{ij} = \partial_i A_j - \partial_j A_i$. 

From this Ward identity it is evident that one can generically violate momentum conservation, while preserving energy density conservation, by introducing background sources in the field theory which depend on the spatial coordinates. (Note that spontaneous breaking of the translational symmetry on its own is not enough to dissipate momentum.) The introduction of such sources is rather natural: a source for $A_i$ with periodicity in the spatial directions represents an ionic lattice while other lattice effects can be captured by a periodic scalar field. 

Holographically, spatially dependent sources for the conserved current can be modelled by a dual gauge field which is spatially modulated. The backreaction of this field onto the metric and other fields gives rise to fields which are stationary but inhomogeneous. In $(d+1)$ bulk dimensions one therefore has to solve partial differential equations in the radial coordinate and the spatial coordinates which are only tractable numerically. Numerical analysis has shown that such 
explicit breaking of translational invariance indeed removes the delta function in the conductivity at zero frequency  \cite{Horowitz:2012ky,Horowitz:2012gs}.

There is considerable interest in the behaviour of the optical conductivity $\sigma(\omega)$ in holographic models at higher frequencies, in the range $T < \omega < \mu$, where $\mu$ is the chemical potential. Over such a range of frequencies certain high temperature superconductors in the normal phase exhibit scaling law behaviour of the form
\be
\sigma(\omega) = K \omega^{\gamma -2} e^{i \frac{\pi}{2} (2 -  \gamma)}
\ee
with $\gamma \approx 1.35 \approx 4/3$ and $K$ a constant. These systems are considered to be strongly coupled with the scaling law potentially a signal of underlying quantum criticality. Rather surprisingly, the introduction of a lattice into holographic models not only results in finite DC conductivity but also apparently induces scaling behaviour in the optical conductivity for a range of frequencies \cite{Horowitz:2012ky,Horowitz:2012gs} (see also \cite{Ling:2013aya,Ling:2013nxa}):
\be
| \sigma| = c + K \omega^{\gamma -2} \label{form}
\ee
with $(c,K)$ constants, $\gamma \approx 1.35$ and the phase of the conductivity approximately constant. Note that $\sigma$ here refers to the homogeneous part of the conductivity. 

Clearly it would be interesting to understand the origin of this scaling behaviour better but the scaling emerges from the numerical analysis and does not make evident which ingredients are crucial to obtain a scaling regime. For example, it is known that one can obtain scaling behaviour for the AC conductivity without explicitly breaking translational invariance; scaling with the correct exponent arises in Einstein-Maxwell-Dilaton models, although solutions with the required value of $\gamma$ appear to be thermodynamically unstable \cite{Gouteraux:2013oca}. While one expects that the scaling is associated with an underlying quantum critical state, the scaling itself emerges at finite temperature and, from the holographic viewpoint, is therefore not associated not only with the spacetime region immediately adjacent to the horizon but also with regions further from the horizon.  From this perspective it is not obvious to what extent the scaling should be sensitive to the details of the far IR or the mechanism of translational symmetry breaking. 

As explored in \cite{Iizuka:2012pn,Donos:2013eha,Donos:2014uba}, simplified models of translational symmetry breaking can be obtained by imposing symmetries on the bulk solutions: one can tune matter field profiles such that the metrics for the equilibrium configurations are homogeneous but anisotropic. The resulting equations of motion therefore simplify, reducing to ordinary differential equations in the radial coordinates, although these equations nonetheless still need to be solved numerically. In such models one does not find scaling behaviour of the AC conductivity, which indicates that this behaviour is non-generic. An interesting feature of these models is that one finds transitions between metallic and insulator behaviour as parameters are adjusted; see also \cite{Donos:2012js,Gouteraux2014,Mefford:2014gia} for related discussions on metal-insulator transitions.

In this paper we will explore the simplest possible models of translational symmetry breaking, namely those for which the inhomogeneous matter field profiles are chosen such that the metrics for the equilibrium configurations remain both homogeneous and isotropic. The equations of motion for the equilibrium black brane solutions can therefore be solved explicitly analytically. The presence of inhomogeneous matter field profiles nonetheless guarantees that momentum can be dissipated by fluctuations propagating around these equilibrium solutions, and therefore one obtains finite DC conductivities. 

Massive gravity models \cite{Vegh:2013sk,Davison:2013jba,Blake:2013owa,Amoretti:2014zha} have been proposed as translational symmetry breaking models of this type. However, massive gravity is
a bottom up phenomenological theory and it is not clear that it is well-defined at the quantum level. The holographic dictionary between the background metric used in massive gravity and the dual field theory is obscure. It is therefore preferable to work with models whose top down origin can be made more manifest.

As discussed above, switching on any operator source with spatial dependence triggers momentum dissipation. Moreover, any scalar field action with shift symmetry admits solutions for which the scalar field is linear in the spatial coordinates and thus the scalar contributions to the stress energy tensor are homogeneous. As shown in 
\cite{Andrade2013}, by choosing an action with a number of massless scalar fields equal to the number of spatial directions one can engineer scalar field profiles such that the bulk stress energy tensor and hence the resulting black brane geometry are both homogeneous and isotropic. See also the earlier work in \cite{Bardoux:2012aw} in which homogeneous and isotropic black branes supported by fluxes were classified; it would be interesting to find AdS/CFT applications for these solutions.

In this paper we will explore general actions with shift symmetry which admit homogeneous and isotropic black brane solutions and realise momentum relaxation. In particular, we will be led to consider square root terms:
\be
{\cal L} = - a_{1/2} \sum_I \sqrt{ (\partial \phi_I)^2} \label{eq:sqr_act}
\ee
where the summation is over spatial directions, labelled by $I$, and reality of the action requires that $\partial \phi_I$ is not timelike. Such Lagrangians clearly have shift symmetry and, as we explain in section \ref{sec:two}, can be used to engineer the required homogeneous and isotropic geometries. 

Square root actions are unconventional but have arisen in several related contexts. For example, time dependent profiles of scalar fields associated with the cuscuton square root action have been proposed in the context of dark energy \cite{Afshordi:2006ad,Afshordi:2007yx}. The same action arose in the context of holography for Ricci flat backgrounds: the holographic fluid on a timelike hypersurface outside a Rindler horizon has properties consistent with a hydrodynamic expansion around a $\phi = t$ background solution of the cuscuton model \cite{Compere:2011dx,Compere:2012mt}. 

We will show in section \ref{sec:two} that the action \eqref{eq:sqr_act} is directly related to one of the mass terms in massive gravity. Four-dimensional massive gravity consists of the usual Einstein-Hilbert term together with mass terms for the graviton $g_{\mu \nu}$  of the following form:
\be
{\cal L} = m^2 \left ( (\alpha_1 \sqrt{g^{\mu\nu} h_{\mu \nu}} + \alpha_2  ( g^{\mu \nu} h_{\mu \nu} - \sqrt{h^{\mu \nu} h_{\mu \nu}}) + \cdots \right ) 
\ee
where $h_{\mu \nu}$ is a reference metric and $h^{\mu \nu} = g^{\mu \rho} g^{\nu \sigma} h_{\rho \sigma}$. The terms in ellipses are higher order in the reference metric and vanish in four dimensions when the reference metric only has two non-vanishing eigenvalues. The coupling constants $\alpha_1$ and $\alpha_2$ are independent. 

It was shown in \cite{Andrade2013} that the $\alpha_2$ term of massive gravity is related to massless scalar fields: the background brane solutions are completely equivalent and certain transport properties (shear modes) agree. Note that not all transport properties agree, since the linearised equations are only equivalent for a subset of fluctuations, those with constrained momenta in the spatial directions. In section \ref{sec:two} we will show that the $\alpha_1$ term of massive gravity is related to the square root terms \eqref{eq:sqr_act}. Again, the background brane solutions are completely equivalent and DC conductivities also agree but as in \cite{Andrade2013} the models are not completely equivalent; even at the linearised level the equations of motion for fluctuations with generic spatial momenta do not agree. The inequivalence between the models is made manifest when one uses a St\"{u}ckelburg formalism for massive gravity.

There has been considerable debate about stability and ghosts in massive gravity, as well as the scale at which non-linear effects occur and effective field theory breaks down, see for example \cite{Boulware:1973my,Deser:2001wx,ArkaniHamed:2002sp,deRham:2010ik,deRham:2010kj,Hassan:2011vm,Hassan:2011hr,Hassan:2011tf}. Clearly all such issues are absent in models based on massless scalar fields but related issues occur in the square root models \eqref{eq:sqr_act}: perturbation theory around the trivial background $\phi_I = 0$ is ill-defined. From the holographic perspective, it is not a priori obvious that the bulk fields $\phi_I$ are dual to local operators in the conformal field theory whose dimensions are real and above the unitary bound and whose norms are positive. 

In section \ref{sec:three} we show that the fields $\phi_I$ are dual to marginal operators in the conformal field theory. The bulk field equations admit a systematic asymptotic expansion near the conformal boundary for any choice of non-normalizable and normalizable modes of these scalar fields, in which all terms in the asymptotic expansion are determined in terms of this data. The bulk action can be holographically renormalized in the standard way. This analysis provides evidence that the action \eqref{eq:sqr_act} is physically reasonable. 

We also show in section \ref{sec:three} that correlations functions of the operators dual to the square root scalar fields $\phi_I$ of \eqref{eq:sqr_act} can be computed in any holographic background in which there are non-vanishing profiles for these fields. These operators  
indeed behave as marginal operators and the norms of their two point functions are positive for $a_{1/2} > 0$. However, the expressions we obtain for the two point functions are not analytic as the background profiles for the scalar fields are switched off. 

The action \eqref{eq:sqr_act} is reminiscent of the volume term in a brane action. In section \ref{sec:three} we show that such actions can indeed arise as tensionless limits of brane actions: the fields $\phi_I$ then correspond to transverse positions of branes. 

\bigskip

In sections \ref{sec:four} and \ref{sec:five} we consider phenomenological models based on massless scalar fields and square root terms:
\begin{equation}
	S = \int  d^{d+1}x \sqrt{-g}\left(R + d(d-1) - \frac{1}{4}F^2 - \sum_{I=1}^{d-1}(a_{1/2}\sqrt{(\del \psi_I)^2} + a_{1}(\del\chi_I)^2)\right)
\end{equation}
Such actions admit charged homogeneous isotropic brane solutions characterised by their temperature, chemical potential and two additional parameters $(\td{\alpha}, \td{\beta})$ associated with the two types of scalar fields $(\psi_I,\chi_I)$ respectively. 

We show that such models have a finite DC conductivity, as expected, and analyse the temperature dependence of the DC conductivity. The parameter $\td{\alpha}$, which is non-zero whenever there are background profiles for the square root fields, leads to a linear increase in the resistivity with temperature at low temperature in a field theory in three spacetime dimensions. In dimensions greater than three the DC conductivity increases with temperature for all values of the parameters $(\td{\alpha}, \td{\beta})$.

We explore the low frequency behaviour of the optical conductivity at low temperature, finding that for all values of our parameters there is a peak at zero frequency, indicating metallic behaviour. However, we show that our models do not fit Drude behaviour even at very low temperature: the effective relaxation constant is complex, indicating that momentum not only dissipates but oscillates. 

Perhaps unsurprisingly, we see no signs of scaling behaviour of the optical conductivity at intermediate frequencies but our numerical analysis indicates minima can arise in the conductivity at intermediate frequencies and low temperatures (in three spacetime dimensions). The behaviour of the optical conductivity in our models is similar to that of heavy fermion compounds: these also have a DC conductivity which increases linearly with temperature at low temperature and they exhibit a transition to a decoherent phase at low temperature in which the conductivity has a minimum at finite frequency. In heavy fermions the origin of this minimum is a hybridisation gap, caused by f-electrons hybridising with conduction electrons, while the dip in the conductivity in our model is a strongly coupling phenomenon, associated with the mixing between scalar and gauge field perturbations. 

\bigskip

The plan of this paper is as follows. In section \ref{sec:two} we explore models for translational symmetry breaking based on shift invariant scalar field actions and we show how such models are related to massive gravity and to scaling limits of branes. In section \ref{sec:three} we analyse square root models, demonstrating that a well-defined holographic dictionary can be constructed. In section \ref{sec:four} we build phenomenological models and compute DC and AC conductivity in these models, showing that features reminiscent of heavy fermions are obtained. In section \ref{sec:five} we analyse generalisations of our models. We conclude in section \ref{sec:six}.

\section{The simplest models of explicit translational symmetry breaking} \label{sec:two}

In this section we consider an Einstein-Maxwell model with cosmological constant, coupled to matter, i.e. an action
\be
S = \int d^{d+1} x \sqrt{-g} \left (R + d (d-1) - \frac{1}{4} F^2  + {\cal L}_{(M)} \right ).
\ee
The gravity and gauge field equations of motion can be written as
\bea
R_{\mu \nu} &=&  - d  g_{\mu \nu}  + \frac{1}{2} (F_{\mu \rho} F_{\nu}^{\; \rho} - \frac{1}{2 (d-1)} F^2 
g_{\mu \nu} ) + \bar{T}_{\mu \nu} ; \\
\nabla_{\mu} ( F^{\mu \nu} ) &=& 0, \nn
\eea
where $\bar{T}_{ \mu \nu}$ is the trace adjusted stress energy tensor for the matter. 

When the matter vanishes, the equations of motion admit the standard electric AdS-RN black brane solution:
\bea
ds^2 &=& \frac{1}{z^2} \left ( - f(z) dt^2 + \frac{dz^2}{f(z)} + dx \cdot dx_{d-1} \right ) \label{ec1} \\
A &=&  \mu (1 - z^{d-2})  dt \nn \\
f &=& (1 - m_0 z^d +  \frac{\mu^2}{\gamma^2} z^{2(d-1)}),
\eea
where $m_{0}$ is the mass parameter and $\mu$ is the chemical potential. It is often convenient to 
choose $m_{0}$ such that 
\be
f(z) = \nn (1 - z^d) + \frac{\mu^2}{\gamma^2} z^d (z^{d-2}  - 1)
\ee
and the horizon is located at $z=1$. The constant $\gamma$ is given by
\be
\gamma^2 = \frac{2(d-1)}{(d-2)}.
\ee
We will consider matter actions which are scalar functionals of the following form:
\be
S_{(M)} =  \int d^{d+1} x \sqrt{-g} {\cal L} (X)
\ee
where $X = (\partial \phi)^2$, i.e. the Lagrangian has shift invariance by construction.
The equation of motion for the scalar in the charged black brane background is then
\be
\nabla^{\mu} \left ( \nabla_{\mu} \phi \frac{\delta {\cal L}}{\delta X} \right ) = 0,
\ee
which, due to the shift symmetry of $X$, always admits the solution 
\be
\phi = {\bf{c}} =  c_{a} x^{a}, \qquad X = z^2 {\bf c \cdot c},
\ee
for any choice of functional of $X$ and any choice of spacelike vector $\bf{c}$\footnote{One could also choose $\phi$ to be linear in time, but such backgrounds would violate energy conservation and will not be considered here.}. The stress energy tensor associated with the scalar matter is given by
\be
T_{\mu \nu} = \frac{1}{2} \left ( - 2 (\partial_\mu \phi)(\partial_\nu \phi) \frac{\delta {\cal L}}{\delta X} + g_{\mu \nu} {\cal L} \right )
\ee
Evaluated on the solution above, this stress energy tensor is by construction homogeneous but not spatially isotropic. 

Now consider a matter action which is a multi-scalar functional of the following form:
\be
S =  \int d^{d+1} x \sqrt{-g} \sum_{I=1}^{d-1}  {\cal L} (X_I) \label{poss1}
\ee
where $X_I = (\partial \phi_I)^2$. The equations of motion in the charged black brane background are
\be
\nabla^{\mu} \left ( \nabla_{\mu} \phi_I\frac{\delta {\cal L}}{\delta X_I} \right ) = 0,
\ee
which admit the solutions
\be
\phi_I = {\bf{c}}_I, \qquad X_I = z^2 {\bf c}_I \cdot \bf{c}_I,
\ee
for any choice of functional and any choices of the spatial vectors ${\bf c}_I = c_{I a} x^a$. The stress energy tensor is
given by
\be
T_{\mu \nu} = \frac{1}{2} \sum_{I=1}^{d-1} \left ( - 2 (\partial_\mu \phi_I)(\partial_\nu \phi_I) \frac{\delta {{\cal L}(X_I)}}{\delta X_I} + g_{\mu \nu} {{\cal L}(X_I)} \right )
\ee
A special case in which spatial isotropy is restored is the following: choose all $(d-1)$ scalar Lagrangians to take the same functional form. Then
by choosing ${\bf c_I} = c x^a$, i.e. $c_{I a} = c$ we obtain a stress energy tensor which restores rotational symmetry in the spatial directions:
\be
T_{\mu \nu} =  \frac{1}{2} \left ( - \sum_{a=1}^{d-1} 2 c^2 \delta_{ab}  \frac{\delta {{\cal L}(X)}}{\delta X} +  (d-1) g_{\mu \nu} {{\cal L}(X)} \right ). \label{upt}
\ee
Here we use the fact that $X_{I}$ evaluated on the solution is $(cz)^2$ for all values of $I$. Therefore for each $I$, both ${\cal L}(X_{I})$ and its derivative take the same values, which we denote without the subscripts. 

Another possibility to restore rotational symmetry in the spatial directions is the following:
\be
S =  \int d^{d+1} x \sqrt{-g} {\cal L} ( \sum_{I=1}^{d-1}  X_I) \label{poss2}
\ee
where $X_I = (\partial \phi_I)^2$. The equations of motion in the charged black brane background remain
\be
\nabla^{\mu} \left ( \nabla_{\mu} \phi_I \frac{\delta {\cal L}}{\delta X_I} \right ) = 0,
\ee
which always admit the solutions
\be
\phi_I= {\bf{c}}_I, \qquad X_I = z^2 {\bf c_I \cdot c_I},
\ee
for any choice of functional and any choices of the spatial vectors ${\bf c}_I = c_{I a} x^a$. The stress energy tensor is
given by
\be
T_{\mu \nu} = \frac{1}{2} \left ( - 2 \sum_{I} (\partial_\mu \phi_I)(\partial_\nu \phi_I) \frac{\delta {{\cal L}(X)}}{\delta X} + g_{\mu \nu} {{\cal L}(X)} \right ),
\ee
where we have defined
\be
X = \sum_{I} X_{I} 
\ee
The special case in which spatial isotropy is restored is the following: choose 
${\bf c_I} = c x^a$, i.e. $c_{I a} = c$
and $X = (d-1) (cz)^2$. The  stress energy tensor is
\be
T_{\mu \nu} =  \frac{1}{2} \left ( - \sum_{a=1}^{d-1} 2 c^2 \delta_{ab}  \frac{\delta {{\cal L}(X)}}{\delta X} +  
g_{\mu \nu} {{\cal L}(X)} \right ),
\ee
which is very similar to the previous form \eqref{upt}.

In summary, given any Lagrangian functional built out of $(d-1)$ scalar fields with shift symmetry, one can construct solutions for which the stress energy tensor preserves spatial isotropy and homogeneity. The backreaction on the black brane metric therefore preserves the usual black brane form for the metric, with a different blackening factor. The breaking of translational invariance by the scalar fields ensures that 
the momenta of fluctuations can be dissipated. In the remainder of this section we will consider the physical interpretations of various types of functionals. 

\subsection{Polynomial Lagrangians}

Consider first the case of \eqref{poss1}. 
If the Lagrangian is of polynomial form, i.e. ${\cal L}(X) = X^m$, then the stress tensor takes the particularly simple form evaluated on the scalar field profiles:
\be
T_{\mu \nu} =  \frac{1}{2} X^m \left ( -  2 m g_{ab}  +  (d-1) g_{\mu \nu} \right )
\ee
where $g_{ab}$ denotes the metric in the spatial directions.
The trace adjusted stress energy tensor $\bar{T}_{\mu \nu}$ is defined 
as
\be
\bar{T}_{\mu \nu} = T_{\mu \nu} - \frac{1}{(d-1)} T g_{\mu \nu}
\ee
and is given by 
\be
\bar{T}_{\mu \nu} =   ( c z )^{2m} \left ( -  m g_{ab} +  (m-1)  g_{\mu \nu} \right ).
\ee
If the Lagrangian can be expressed as a sum of such terms, namely ${\cal L}(X) =  - \sum_{m} a_{m} X^{m}$, the corresponding trace adjusted stress energy tensor is 
\be
\bar{T}_{\mu \nu} =    \sum_{m} a_m ( c_m z )^{2m} \left (  m g_{ab}  -  (m-1)  g_{\mu \nu} \right ). \label{gen1}
\ee
Note that this class includes the special case of $m=1$, i.e. massless scalar fields. 

Scalar field profiles for which the stress energy tensor preserves rotational symmetry in the spatial directions by construction give rise to backreacted solutions which much satisfy a homogeneous black brane metric ansatz 
\bea
ds^2 &=& \frac{1}{z^2} \left ( - F(z) dt^2 + \frac{dz^2}{F(z)} + dx \cdot dx_{d-1} \right ) \label{ec2} \\
A &=&  \mu (1 - z^{d-2}) dt.  \nn 
\eea
In the limit that the matter fields vanish $F(z)$ coincides with the $f(z)$ given in the previous section, \eqref{ec1}. Using the Ricci tensor for the metric \eqref{ec2},
\bea
R_{tt} &=& \left ( - d F + \frac{1}{2} (d+1) z F' - \frac{1}{2} z^2 F'' \right ) g_{tt};  \\
R_{zz} &=& \left (- d F  + \frac{1}{2} (d+1) z F' - \frac{1}{2} z^2 F'' \right ) g_{zz}; \nn \\
R_{ab} &=& \left ( - d F + z F' \right) g_{ab}, \nn
\eea
we note that such an ansatz is required given the form of the matter stress energy tensor. 

The solution for the blackening function $F$ can be written as 
\be
F(z) = f(z) + \sum_{m} \frac{a_m}{2m -d} ( c_m z )^{2m},
\ee
with $f(z)$ given previously in \eqref{ec2}. This expression assumes that $d \neq 2m$; in the latter case the solution for $F$ involves logarithms, i.e. we obtain a term
\be
a_{m} (c_{m} z)^d \ln (z),
\ee
which gives rise to non-analytic behaviour. 

Solutions to \eqref{poss2} in the case that ${\cal L}$ is polynomial, i.e. ${\cal L} = - b_m (\sum_{\lambda} X_{\lambda})^m$, are very similar. Evaluated on the scalar field profiles one obtains
\be
T_{\mu \nu} =  - \frac{b_m}{2} X^m \left ( -  \frac{2 m}{(d-1)} g_{ab}  +  g_{\mu \nu} \right ).
\ee
with $X = (d-1) c_m^2 z^2$. Therefore the trace adjusted stress energy tensor is 
\be
\bar{T}_{\mu \nu} = - b_m \left ( (d-1) c_m^2 z^2) \right )^m \frac{1}{(d-1)} \left ( - mg_{ab} + (m-1) g_{\mu \nu} \right )
\ee
This coincides with the expression above in the case of $m=1$ (massless scalar fields) as the Lagrangians are the same. The corresponding solutions for the blackening functions are
\be
F(z) = f(z) +  \frac{b_m}{2m -d} (d-1)^{m-1} ( c_m z )^{2m},
\ee
with $f(z)$ given previously in \eqref{ec2}. Again the case $d = 2m$ will involve logarithmic terms.

\subsection{Relation to massive gravity}

In this section we will discuss the relation between massive gravity and our scalar field models. Let us consider the following Lagrangian
\be
{\cal L} = - a_{1/2} \sum_{I} \sqrt{ (\partial \phi_I)^2}  - a_{1} \sum_{I} (\partial \chi_I)^2. \label{4d}
\ee
The  trace adjusted stress energy tensor is 
\bea
\bar{T}_{\mu \nu} &=& \frac{1}{2 (d-1)} a_{1/2}  \sum_{I} \sqrt{ (\partial \phi_I)^2} g_{\mu \nu}
+ \frac{1}{2} a_{1/2} \sum_{I} \frac{1}{  \sqrt{ (\partial \phi_I)^2}} \partial_{\mu} \phi_{I} \partial_{\nu} \phi_{I} \\
&& + a_1 \sum_{I} \partial_{\mu} \chi_I \partial_{\nu} \chi_I \nn
\eea
The scalar field profiles 
\be
\phi_I = c_{1/2} x^{I}; \qquad
\chi_I = c_1 x^{I}
\ee
give rise to a trace adjusted stress energy tensor which is
\be
\bar{T}_{\mu \nu} =    \frac{1}{2} a_{1/2} ( c_{1/2} z ) \left (  g_{ab}  +    g_{\mu \nu} \right ) 
+ a_1 (c_1 z)^2 g_{ab}. \label{o1}
\ee
The backreacted blackening function is
\be
F(z) = f(z) - \frac{1}{(d-1)} a_{1/2} c_{1/2} z - \frac{1}{(d-2)}a_1 (c_1 z)^2,
\ee
which in $d=3$ has precisely the same form as the massive gravity solution found in \cite{Vegh:2013sk}. 

One can also consider a slightly different Lagrangian
\be
{\cal L} = - a_{1/2} \left ( \sqrt{ \sum_I (\partial \phi_I)^2} \right ) - a_{1} \left ( \sum_I (\partial \chi_I)^2  \right ). \label{4d2}
\ee
for which the trace adjusted stress energy tensor is 
\bea
\bar{T}_{\mu \nu} &=&  \frac{1}{2 (d-1)} a_{1/2} \sqrt{\sum_{I}  (\partial \phi_I)^2} g_{\mu \nu}  + 
\frac{1}{2} a_{1/2}  \frac{1}{  \sqrt{ \sum_{I} (\partial \phi_I)^2}} \sum_{I} \partial_{\mu} \phi_{I} \partial_{\nu} \phi_{I} \label{o22}
\\
&& +  a_1 \sum_{I} \partial_{\mu} \chi_I \partial_{\nu} \chi_I. \nn
\eea
Evaluated on the scalar field profiles
\be
\phi_{I} = c_{1/2} x^{I}; \qquad
\chi_{I} = c_1 x^{I}
\ee
the trace adjusted stress energy tensor  becomes
\be
\bar{T}_{\mu \nu} =    \frac{1}{2 \sqrt{d-1} } a_{1/2} ( c_{1/2} z ) \left (  g_{ab}  +    g_{\mu \nu} \right ) 
+ a_1 (c_1 z)^2 g_{ab}. \label{o2}
\ee
The corresponding backreacted blackening function is
\be
F(z) = f(z) - \frac{1}{(d-1)^{\frac{3}{2}}} a_{1/2} c_{1/2} z - \frac{1}{(d-2)}a_1 (c_1 z)^2,
\ee
which in $d=3$ again has precisely the same form as the massive gravity solution found in \cite{Vegh:2013sk} and further analysed in 
\cite{Davison:2013jba} and \cite{Blake2013}.

To understand the relation with massive gravity in four bulk dimensions, let us first recall that the action for massive gravity consists of the Einstein-Hilbert terms plus the following mass terms:
\be
{\cal L} = m^2 \sum_{i} \alpha_{i} {\cal U}_{i}(g,h),
\ee
where in terms of the matrix ${\cal K}^{\mu}_{\nu} \equiv \sqrt{ g^{\mu \rho} h_{\rho v}}$
\bea
{\cal U}_1 &=& \left [ {\cal K} \right ]; \label{four} \\
{\cal U}_2 &=& \left [ {\cal K} \right ]^2 - \left [ {\cal K}^2 \right ]; \nn \\
{\cal U}_3 &=& \left [ {\cal K} \right ]^3 - 3  \left [ {\cal K} \right ] \left [ {\cal K}^2 \right ] + 2 \left [ {\cal K}^3 \right ]; \nn \\
{\cal U}_4 &=& \left [ {\cal K} \right ]^4 - 6 \left [ {\cal K}^2 \right ] \left [ {\cal K} \right ]^2 +8  \left [ {\cal K}^3 \right ] \left [ {\cal K} \right ] - 3 \left [ {\cal K}^2 \right ]^2 -6  \left [ {\cal K}^4 \right ]; \nn
\eea
The notation $\left [ Y \right ]$ denotes the matrix trace. 
Here $h_{\mu \nu}$ is a reference metric, which can be expressed via a coordinate transformation in terms of scalar (St\"{u}ckelburg) fields $\pi^a$ whenever it is flat, i.e. 
\be
h_{\mu \nu} = \eta_{ab} \partial_{\mu} \pi^a \partial_{\nu} \pi^b
\ee
Unitary gauge is then defined as $\pi^a = x^{\mu} \delta_{\mu}^{a}$. The restriction to a degenerate reference metric in which only the spatial components are non-vanishing was obtained in \cite{Vegh:2013sk} using only two non-vanishing scalar fields, $\pi^1$ and $\pi^2$, which take an analogous form to those given above, namely
\be
\pi^1 = x^1; \qquad \pi^2 = x^2 \label{fr}
\ee
From the first two terms in \eqref{four} one obtains a trace adjusted stress energy tensor
\be
\bar{T}_{\mu \nu} = \frac{1}{2} m^2 \alpha_1 ({\cal K}_{\mu \nu} + \frac{1}{2} \left [ {\cal K} \right ] g_{\mu \nu}) 
 - m^2  \alpha_2 ({\cal K}^2_{\mu \nu} -  \left [ {\cal K} \right ] {\cal K}_{\mu \nu}). \label{Se1}
\ee
The final two terms in \eqref{four} give rise to a vanishing stress energy tensor in four dimensions, as expected, as the spatial gauge only involves two non-vanishing eigenvalues for the matrix. Higher order terms in $[{\cal K} ]$ would contribute in dimensions greater than four but massive gravity in dimensions higher than four has not been explored in detail in earlier literature.

Evaluated on the particular background given by the two scalar fields
\bea
\bar{T}_{\mu \nu} = \frac{1}{2} m^2 \alpha_1 z ({g}_{ab} + g_{\mu \nu}) 
 + m^2  \alpha_2 z^2 g_{ab},  \label{mg1}
\eea
where we assume that the metric ansatz $g_{11} = g_{22} = z^{-2}$ remains consistent, which is then justified a posteriori.  The expressions \eqref{mg1} and \eqref{o1} clearly match under the identifications
\be
m^2 \alpha_1 = a_{1/2} c_{1/2}; \qquad
m^2 \alpha_2 = a_1 c_1^2, \label{eq:mg}
\ee
and \eqref{mg1} and \eqref{o1} similarly can be matched. 

While the black brane solutions in our models match those of massive gravity, it is clear that fluctuations and hence transport properties of these solutions will differ between massive gravity and the scalar field models. For the terms quadratic in ${\cal K}$, corresponding to the massless scalar fields in our models, this issue was discussed in \cite{Andrade2013}. Focussing on the terms linear in ${\cal K}$, note that 
\be
[ {\cal K} ] = \sqrt{ (\partial \pi^1)^2 + (\partial \pi^2)^2}, 
\ee
and therefore the second term in \eqref{Se1} seems to ressemble the first term in \eqref{o22}. However, the scalar fields in the massive gravity model are assumed to depend only on the spatial components as given in \eqref{fr} whereas the scalar fields in \eqref{o22} are completely unrestricted. The first term in \eqref{Se1} can be written explicitly in terms of 
\be
{\cal K}_{\mu \nu} = g_{\mu \rho} \sqrt{ \sum (\partial^{\rho} \pi) (\partial_{\nu} \pi) },
\ee
which is not of the same form as the second term in \eqref{o22} unless we restrict the scalar fields to the form \eqref{fr}. In the background brane solutions the scalar fields indeed necessarily take the form \eqref{fr} but this property cannot generically hold for fluctuations around the equilibrium solution. We will show in section \ref{sec:four} why the conductivities nonetheless match those of massive gravity. 

Another conceptual difference between our model and massive gravity is the following. In our models the scalar fields $\phi_{I}$ and $\chi_{I}$ are treated as independent fields but in massive gravity they are identified 
as the same field. As we discuss in section \ref{sec:five}, it is however straightforward to restrict to the case in which these fields are identified. 

\subsection{Relation to branes}

From the perspective of top-down models, the appearance of square root terms is unconventional. 
In this section we will show that similar terms can arise from tensionless limits of branes. 
Consider the following action:
\be
S = - b_{1/2}\int d^{d+1} x   \sqrt{ - \det (g_{\mu \nu} + \sum_{I =1}^{d-1} \partial_{\mu} \phi_{I} \partial_{\nu} \phi_{I}) } \equiv  - b_{1/2} \int d^{d+1} x  \sqrt{- \det{M}}. \label{eq:brane1}
\ee
This action can be interpreted in terms of a brane with a $(d+1)$ dimensional world volume, which is probing $(d-1)$ flat transverse directions. To show this, recall that the DBI term in the action for a $p$-brane is
\be
S_{DBI} = - T  \int d^{p+1} x \sqrt{ - \det (g_{MN} \partial_{\mu} X^{M} \partial_{\nu} X^{N} +  F_{\mu \nu})},
\ee
where $T$ is the brane tension, $F_{\mu \nu}$ is the worldvolume gauge field strength; $X^{M}$ are the brane positions and $g_{MN}$ is the metric of the spacetime in which the brane propagates. Fixing static gauge for the brane corresponds to choosing $X^{\mu} \equiv x^{\mu}$ and the gauge fixed action is
\be
S_{DBI} = - T \int d^{p+1} x \sqrt{- \det (g_{\mu \nu} + g_{m ( \mu} \partial_{\nu )} X^{m} + g_{mn} \partial_{\mu} X^{m} \partial_{\nu} X^{n} + F_{\mu \nu})},
\ee
where the transverse coordinates are denoted as $X^{n}$. Whenever the background metric is diagonal $g_{m \nu} = 0$. 
If there are no Wess-Zumino terms sourcing the gauge field, then the gauge field strength may also always be set to zero. This results in a brane probing the transverse directions:
\be
S_{DBI} = - T \int d^{p+1} x \sqrt{- \det(g_{\mu \nu} + g_{mn} \partial_{\mu} X^{m} \partial_{\nu} X^n)},
\ee
and clearly when $g_{mn} = \delta_{mn}$ the action reduces to \eqref{eq:brane1}, with $b_{1/2}$ being identified as the brane tension. 

The trace adjusted stress energy tensor following from \eqref{eq:brane1} is 
\be
\bar{T}_{\mu \nu} = - \frac{b_{1/2} \sqrt{- \det {M}}}{2 \sqrt{-g}} 
\left ( M_{\mu \nu} - \frac{1}{(d-1)} M^{\rho \sigma} M_{\rho \sigma} g_{\mu \nu} \right ). 
\ee
Again the specific solution $\phi_{I} = c x^{I}$ preserves spatial homogeneity and isotropy with the
trace adjusted stress energy tensor being 
\be
\bar{T}_{\mu \nu} =  - \frac{1}{2} b_{1/2} (1 + c^2 z^2)^{\frac{d-1}{2}} \left ( \frac{2}{(1-d)} g_{\mu \nu} - c^2 z^2 g_{\mu \nu} + c^2 \delta_{ab} \right ) 
\ee
and the blackening factor taking the form
\be
F(z) = f(z) +  b_{1/2} z^d \int 
\frac{dz}{(d-1) z^{d+1}} (1 + c^2 z^2)^{\frac{d-1}{2}}. \label{brane1}
\ee
Expanding the second term for small $z$ near the AdS boundary gives
\be
 - b_{1/2} \left ( \frac{1}{d(d-1)} + \frac{c^2 z^2}{2 (d-2)} + \cdots \right), 
\ee
i.e. there is an effective shift of the AdS radius as well as subleading terms in the expansion. When $d$ is odd the integral gives an analytic expression; for example, for $d=3$ one obtains 
\be
F(z) = f(z) - b_{1/2} \left ( \frac{1}{6} + \frac{1}{2} c^2 z^2 \right )
\ee
but $d$ even generates logarithmic terms and therefore $F(z)$ is not analytic, e.g. for $d=4$ 
\be
F(z) = f(z) - b_{1/2} \left ( \frac{1}{8} + \frac{3}{4} c^2 z^2 - \frac{3}{2} c^4 z^4 {\rm{ln}}(z) - \frac{1}{4} c^6 z^6. \right )
\ee
Working perturbatively around AdS, this brane type Lagrangian leads to a shift in the cosmological constant along with a spectrum of $(d-1)$ massless scalar fields, dual to $(d-1)$ marginal couplings in the field theory. It therefore reproduces analogous behaviour to the massless scalar fields discussed in the previous sections. 


\bigskip

Another brane model can be obtained as follows. Consider $(d-1)$ branes of equal tension, each probing one transverse flat direction only: 
\be \label{eq:brane_act}
S = - b_{1/2} \sum_{I=1}^{d-1} \int d^{d+1} x   \sqrt{ - \det (g_{\mu \nu} + \partial_{\mu} \phi_{I} \partial_{\nu} \phi_{I}) }. 
\ee
Using Sylvester's determinant theorem this action can be rewritten as
\be
S = - b_{1/2} \sum_{I=1}^{d-1} \int d^{d+1} x \sqrt{- \det{g}} \sqrt{1 + \partial^{\mu} \phi_I \partial_{\mu} \phi_I}. \label{brane2}
\ee
The trace adjusted stress tensor is
\be
\bar{T}_{\mu \nu} = - \frac{b_{1/2} \sum_{I} \sqrt{- \det {M_I}}}{2 \sqrt{-g}} 
\left ( M_{I \mu \nu} - \frac{1}{(d-1)} M_I^{\rho \sigma} M_{I \rho \sigma} g_{\mu \nu} \right )
\ee
where now
\be
M_{I \mu \nu} = g_{\mu \nu} + \partial_{\mu} \phi_{I} \partial_{\nu} \phi_{I}.
\ee
The solution with $\phi_I = c x^I$ is homogeneous and isotropic with the blackening factor being
\be
F(z) = f(z) +  b_{1/2} z^d \int 
\frac{dz}{ z^{d+1}} (1 + c^2 z^2)^{\frac{1}{2}} \left ( 1 + \frac{1}{2} c^2 z^2 (d-2) \right ),
\ee
which coincides with \eqref{brane1} in $d=2$. 

The action \eqref{brane2} admits a scaling limit in which the brane tension is taken to zero $b_{1/2} \rightarrow 0$ with $\psi_I = b_{1/2} \phi_I$ remaining finite. This limit results in 
\be
S \approx - \sum_{I=1}^{d-1} \int d^{d+1} x \sqrt{-g} \sqrt{ (\partial \psi_I)^2},
\ee
which is of the square root form. 

\section{Square root models} \label{sec:three}

While one can obtain solutions for any polynomial functional, one would usually restrict to the case of $m=1$, i.e. massless scalar fields.  In AdS/CFT the operators dual to these scalar fields are marginal scalar operators and the bulk scalar profiles are therefore immediately interpretable in the dual theory as linear profiles for the associated couplings. 

For integer $m > 1$ the action is higher derivative and for non-integer $m$ the action would be considered non-local. In this section we will argue that both cases may in some limits nonetheless be relevant in bottom up models.

Consider first the case of integer $m > 1$. In the previous section, we assumed that the scalar fields appearing in polynomials of different order were independent. However, in the solutions of interest, the scalar field profiles are the same for each order polynomial. Therefore there is no reason why we should not identify the scalar fields, e.g. we could consider
\be
{\cal L} = - a_1 \sum_{I} (\partial \phi_I)^2  - a_2 \sum_{I} ( (\partial \phi_I)^2)^2
\ee
or 
\be
{\cal L} = - a_1 \sum_{I} (\partial \phi_I)^2  - a_2 (\sum_{I} (\partial \phi_I)^2 )^2
\ee
The fourth order terms can be viewed as higher derivative corrections to the leading order action; $a_2$ should therefore be considered as parametrically small compared to $a_1$ (which can always be rescaled to the canonical value by rescaling the fields). However, it makes sense to consider how a small $a_2$ would affect transport and thermodynamic properties of charged black branes, although we will not pursue this further here.

Now let us turn to non-integer $m$, focussing on the case of $m =1/2$, i.e. a Lagrangian of the form
\be
{\cal L} = - \sqrt{ (\partial \phi)^2} \label{sqrt}
\ee
Note that reality of the Lagrangian requires that $(\partial \phi)$ is not timelike. This is certainly an unconventional Lagrangian in holography, although similar actions have arisen in several contexts. For example, time dependent profiles of scalar fields associated with the cuscuton action
\be
{\cal L} = \sqrt{ - (\partial \phi)^2}
\ee
have been proposed in the context of dark energy \cite{Afshordi:2006ad,Afshordi:2007yx}. The same action arose in the context of holography for Ricci flat backgrounds: the holographic fluid on a timelike hypersurface outside a Rindler horizon has properties consistent with a hydrodynamic expansion around a $\phi = t$ background solution of the cuscuton model \cite{Compere:2011dx,Compere:2012mt}. 

In the remainder of this section we will explore the behaviour of the \eqref{sqrt} model. 
It is  subtle to work at linear order around an AdS background as the corresponding scalar field equations remain non-linear  in this limit:  we obtain a field equation of the form
\be
\bar{\nabla}^{\mu} \left ( \frac{1}{\sqrt{ (\partial \phi)^2}}  \bar{\nabla}_{\mu} \phi  \right ) = 0, \label{lin-eq}
\ee
where $\bar{\nabla}_{\mu}$ is the AdS connection and here $\phi$ is implicitly treated perturbatively, i.e. the amplitude of the scalar field is small. 

When one works perturbatively around the AdS background, we need to take into account the fact that the scalar field perturbation is of the same order as the backreaction of the metric. (Note that in the exact, non-linear, black brane solutions  the backreaction on the metric is indeed of the same order as the scalar field itself.) 
It is convenient to express the coupled metric and scalar field equations using
\bea
\nabla_{\mu} v^{\mu} &=& 0; \label{vel1} \\
\bar{T}_{\mu \nu} &=& - \frac{1}{2} \sqrt{X} ( g_{\mu \nu} + 2 v_{\mu} v_{\nu} ), \nn
\eea
where $X = \sqrt{(\partial \phi)^2}$ while the velocity field $v_{\mu}$ is conserved and satisfies $v^{\mu} v_{\mu} = 1$. In terms of the scalar field one can express the velocity field as the gradient flow
\be
v_{\mu} = \frac{\nabla_{\mu} \phi}{\sqrt{X}}. \label{vel2}
\ee
Working perturbatively around the AdS background requires that $\phi \sim \delta$ with $\delta \ll 1$. The metric perturbation is then of the same order as the scalar field and the non-linearity is manifest in the fact that the velocity field is of order one. 

We can now proceed to solve these equations as follows. Working perturbatively in the amplitude $\delta$ let
\be
v^{\mu} = v_{0}^{\mu} + v_{1}^{\mu} \delta + {\cal O}(\delta^2)
\qquad
g_{\mu \nu} = \bar{g}_{ \mu \nu} + h_{1 \mu \nu} \delta + {\cal O}(\delta^2),
\ee
where $\bar{g}_{\mu \nu}$ is the AdS metric and $v_0$ is any conserved globally spacelike vector in this metric, which can then be normalised such that $v_0^{\mu} v_{0 \mu} = 1$. Solving the conservation equation up to order $\delta$ gives
\be
v_{1 \mu} = - \frac{1}{2} v_{0 \mu} h_1
\ee
with $h_1 = \bar{g}^{\mu \nu} h_{1 \mu \nu}$. Substituting into the trace adjusted stress energy tensor we
obtain at order $\delta$
\be
\bar{T}_{1 \mu \nu} = - 3 h_{1 \mu \nu} - \frac{1}{2} \sqrt{X} (\bar{g}_{ \mu \nu} + 2 v_{0 \mu} v_{0 \nu}),
\ee
with $\sqrt{X}$ a function of the spacetime coordinates. Therefore the metric perturbation $h_{1 \mu\nu}$ is determined by the Einstein equation in terms of $\sqrt{X}$ and the conserved vector field $v_{0 \mu}$.  Using the linearised Ricci tensor in de Donder gauge ($\bar{\nabla}^{\mu} h_{1 \mu \nu} = 0$) gives
\be
\frac{1}{2} \bar{\Box} h_{1\mu \nu} + \frac{1}{2} \bar{\nabla}_{\mu} \bar{\nabla}_{\nu} h_1 + h_{1} \bar{g}_{\mu \nu} =    \frac{1}{2} \sqrt{X} (\bar{g}_{ \mu \nu} + 2 v_{0 \mu} v_{0 \nu}).
\ee
Tracing this equation with $\bar{g}^{\mu \nu}$ results in 
\be
\bar{\Box} h_1 + (d+1) h_1 = (d+ 3) \sqrt{X}.
\ee
Therefore, the leading order defining data is a scalar field satisfying \eqref{lin-eq}, which is a non-linear equation; we will discuss its solution in section \ref{2pf}. 
Note that an asymptotic expansion of the field equations near the conformal boundary exists as we will discuss in the next section \ref{hol-ren}.

As in the cuscuton and holographic fluid models, one can find simple solutions of the equations of motion with non-vanishing scalar field profiles, and the equations of motion are linear when expanded around such backgrounds. To understand this, let us consider the action
\be
{\cal L} = - a_{1/2} \sum_{I} \sqrt{  (\partial \phi_{I})^2} \label{pot}
\ee
for which the coupled gravity/scalar field system admits the homogeneous and isotropic solution 
\be
ds^2 =  \frac{1}{z^2} \left (   - (1 - \frac{a_{1/2} c_{1/2}}{(d-1)} z)  dt^2 + \frac{dz^2}{(1 - \frac{a_{1/2} c_{1/2}}{(d-1)} z)} + dx \cdot dx_{d-1} \right ) \label{bhsol}
\ee
with $\phi_{I} = c_{1/2} \delta_{I a} x^a$. This is the $\mu \rightarrow 0$, $m_0 \rightarrow 0$ limit of the solution given in the previous section. As discussed above, the backreaction on the metric is linear in the scalar field amplitude and therefore cannot be neglected even for small $c_{1/2}$. 

Note that when $a_{1/2} > 0$ the geometry has a horizon, with the entropy and temperature being
\be
{\cal S} = \frac{1}{4 G_{d+1}} V_{d-1} \left ( \frac{a_{1/2} c_{1/2}}{(d-1)} \right )^{d-1}; \qquad T = \frac{a_{1/2} c_{1/2}}{4 \pi (d-1)},
\ee
where $G_{d+1}$ is the Newton constant.
Analogous behaviour was noted in the massless scalar field model of \cite{Andrade2013}. If $a_{1/2} < 0$ there is a curvature singularity as $z \rightarrow \infty$; this can most easily be seen from the expression for the Ricci scalar
\be
R = \left  ( - d(d+1) F - zz^2 F'' + 2 d z F' \right ) = \left (-d(d+1) + d a_{1/2} c_{1/2} z  \right ).
\ee
The singularity is at infinite proper distance and would presumably therefore not affect the computation of correlation functions. It is a good singularity, in the sense of \cite{Gubser:2000nd}, since it is shielded in the black brane solutions of the previous section for which $m_0 > 0$.  We will discuss the linearised equations of motion in such backgrounds in section \eqref{2pf}.

\subsection{Holographic renormalisation for square root models} \label{hol-ren}

In this section we explore asymptotically locally AdS solutions of the action
\be
S = \frac{1}{16 \pi G_{d+1}} \int d^{d+1} x \sqrt{-g} \left ( R + d (d-1) - a_{1/2} \sum_{I}^{d-1} \sqrt{ (\partial \phi_I)^2} \right ).
\ee
Despite the subtleties discussed in the previous section, one can solve the field equations iteratively near the conformal boundary and systematically set up holographic renormalisation in the standard way \cite{Henningson:1998gx,Balasubramanian:1999re,deHaro:2000xn}.

The onshell action, including Gibbons-Hawking boundary term, is
\be
S_{{\rm bare}} = \frac{1}{16 \pi G_{d+1}} \int d^{d+1} x \sqrt{-g} \left (-2 d  + a_{1/2} \sum_{I}^{d-1} \sqrt{ (\partial \phi_I)^2} \right ) - \frac{1}{8 \pi G_{d+1}} \int d^d x \sqrt{- \gamma} K. \label{act1}
\ee
One can rewrite the bulk scalar field term as
\bea
&&  \frac{1}{16 \pi G_{d+1}} \int d^{d+1} x \sqrt{-g} \left ( a_{1/2} \sum_{I}^{d-1} \frac{ (\partial \phi_I)^2}{\sqrt{ (\partial \phi_I)^2}} \right ) \\
&& = 
  \frac{1}{16 \pi G_{d+1}} \int d^{d+1} x \sqrt{-g}  a_{1/2} \sum_{I}^{d-1}  \left ( \nabla_{\mu} \left ( \frac{ \phi_I \partial^{\mu} \phi_I}{\sqrt{ (\partial \phi_I)^2}} \right ) - \phi_I \nabla_{\mu} \left ( \frac{\partial^{\mu} \phi_I}{\sqrt{ (\partial \phi_I)^2}} \right ) \right ) \nn \\
  && =  \frac{1}{16 \pi G_{d+1}} \int d \Sigma_{\mu}  a_{1/2} \sum_{I}^{d-1}   \left ( \frac{ \phi_I \partial^{\mu} \phi_I}{\sqrt{ (\partial \phi_I)^2}} \right ), \nn
\eea
where we use the scalar field equation in the final equality.

In the neighbourhood of the conformal boundary the metric can be expanded as
\be
ds^2 = \frac{d \rho^2}{4 \rho^2} + \frac{1}{\rho} g_{ij} dx^i dx^j
\ee
where 
\be
g_{ij} = g_{(0) ij} (x) + \rho^{1/2} g_{(1) ij} (x) + \rho  g_{(2) ij} (x) + \cdots + \rho^{\frac{d}{2}} \left ( g_{(d) ij}(x) + \ln (\rho) h_{(d) ij} \right ) + \cdots  \label{m1}
\ee
We will show that there exist scalar field solutions such that 
\be
\phi_I = \phi_{(0)I} (x) + \rho \phi_{(2)I} (x) + \cdots + \rho^{\frac{(d+1)}{2}} (\phi_{(d+1)/2 I}(x) + \ln (\rho) \td{\phi}_{(d+1)/2 I}(x)) + \cdots \label{s1}
\ee
and the metric expansion takes the above form. 

In the Fefferman-Graham coordinate system the Ricci tensor can be expressed as
\bea
R_{\rho \rho} &=&  \frac{1}{4} {\rm Tr} (g^{-1} g')^2 - \frac{1}{2} {\rm Tr} (g^{-1} g'') - \frac{d}{4 \rho^2}; \\
R_{\rho j} &=& \frac{1}{2} \nabla^i g'_{ij} - \frac{1}{2} \nabla_j ( {\rm Tr} (g^{-1} g') ); \nn \\
R_{ij} &=&  {\cal R}_{ij} + (d-2) g'_{ij} + {\rm Tr}(g^{-1} g') g_{ij} - \rho (2 g'' - 2 g' g^{-1} g' + {\rm Tr} (g^{-1} g') g')_{ij} - d \frac{g_{ij}}{\rho}, \nn
\eea
where ${\cal R}$ is the curvature of $g_{ij}$, for which the associated connection is $\nabla_i$. 

The scalar field equations can be expressed in this coordinate system as
\be
\rho^{1 + \frac{d}{2}} \frac{1}{\sqrt{-g}} \partial_{\rho} \left ( \frac{4 \sqrt{-g}}{\rho^{\frac{d-1}{2}} } \frac{\partial_{\rho} \phi_I}{Y_I} \right ) 
+ \rho^{\frac{1}{2}}  \nabla_{i} \left ( \frac{\partial^{i} \phi_I}{Y_I} \right ) = 0, \label{sca1}
\ee
where implicitly $\partial^i \phi_I = g^{ij} \partial_{j} \phi_I$ and
\be
Y_I = \sqrt{ g^{jk} \partial_j \phi_I \partial_k  \phi_I + 4 \rho (\partial_{\rho} \phi_I)^2 }
\ee
The trace adjusted stress energy tensor can be written as
\bea
\bar{T}_{\rho \rho} &=& - \frac{d}{4 \rho^2}  + \frac{a_{1/2} }{2 \rho^{3/2}} \sum_{I} \frac{1}{Y_I } \left (\rho  \partial_{\rho} \phi_I \partial_\rho \phi_I + \frac{1}{4 (d-1) } (g^{ij} \partial_i \phi_I \partial_j \phi_I ) + \frac{\rho }{(d-1)} (\partial_\rho \phi_I)^2  \right ); \nn \\
\bar{T}_{\rho i} &=&   \frac{a_{1/2}}{2 \rho^{1/2}}  \sum_{I} \frac{1}{Y_I} \left ( \partial_{i} \phi_I \partial_\rho \phi_I \right );  \\
\bar{T}_{ij} &=& -  \frac{ dg_{ij}}{\rho} +  \frac{a_{1/2}}{2 \rho^{1/2}}  \sum_{I} \frac{1}{Y_I} \left ( \partial_{i} \phi_I \partial_j \phi_I + \frac{1}{(d-1)} (g^{kl} \partial_k \phi_I \partial_l \phi_I + 4 \rho (\partial_{\rho} \phi_I)^2 ) g_{ij } \right ). \nn
\eea
Tracing the $(ij)$ Einstein equations with $g^{ij}$ gives 
\bea
&& {\cal R} + 2 (d-1) {\rm Tr} (g^{-1} g') - 2 \rho {\rm Tr} (g^{-1} g'') + 2 \rho {\rm Tr} (g^{-1} g')^2 - \rho ( {\rm Tr} (g^{-1} g') )^2 \label{tr-ric} \\
&& \qquad = \frac{1}{2 \rho^{1/2}} a_{1/2} \sum_{I} \left ( \frac{\partial^i \phi \partial_i \phi}{Y_I} + \frac{d}{(d-1)} Y_I \right) . \nn
\eea
The latter equation is not independent but is useful in the analysis below. 

The leading order term in the scalar field equation is at order $\rho^{1/2}$ and enforces
\be
\phi_{(2)I} = \frac { \sqrt{g_{(0)}^{kl} \partial_k \phi_{(0)I} \partial_l \phi_{(0)I}}}{2 (d-1)} \nabla_{(0) i} \left ( \frac{\partial^i \phi_{(0) I}}{\sqrt{g_{(0)}^{kl} \partial_k \phi_{(0)I} \partial_l \phi_{(0)I}}} \right ),
\ee
where all indices are raised using $g^{(0) ij}$ and $\nabla_{(0)ij}$ is the connection of $g_{(0) ij}$. This expression may be written more compactly using the shorthand notation of $Y_{(0) I}$ for the square root term as
\be
\phi_{(2) I} = \frac{Y_{(0)I}}{2 (d-1)}  \nabla_{(0) i} \left (  \frac{\partial^i \phi_{(0) I}}{Y_{(0)I} }  \right ) .
\ee
Using the radial terms in \eqref{sca1} one can see that the normalizable mode of the scalar field occurs at order $(d+1)/2$ in the expansion. The coefficient of this term, $\phi_{(d+1)/2 I}(x)$, is undetermined by the asymptotic analysis. In general one also needs a logarithmic term at the same order to satisfy the field equation; this term $\td{\phi}_{(d+1)/2 I}(x)$ is determined in terms of $\phi_{(0) I}(x)$, as we will see below. 

From the leading $\rho^{-3/2}$ component of the $(\rho \rho)$ Einstein equation one obtains
\be
{\rm Tr} (g_{(0)}^{-1} g_{(1)}) = a_{1/2} \sum_{I} \frac{1}{(d-1)} Y_{(0)I}
\ee
From the leading $\rho^{-1/2}$ component of the $(ij)$ Einstein equations one finds
\be
g_{(1)ij} = a_{1/2} \sum_{I} \frac{1}{ (d-1) Y_{(0) I} } \partial_i \phi_{(0) I} \partial_{j} \phi_{(0) I}
\ee
which is manifestly consistent with the trace. This equation is also consistent with the exact solution \eqref{bhsol} expressed as a Fefferman-Graham expansion. 

The expansion up to this order is sufficient to determine the counterterms
\be
S_{ct} = - \frac{1}{16 \pi G_{d+1}} \int d^{d} x \sqrt{-\gamma} \left ( 2 (1- d) + a_{1/2} \frac{1}{(d-1)^2} \sum_{I}^{d-1} \sqrt{ (\partial \phi_I)^2}+ \cdots \right ), \label{act2}
\ee
where the first term is the standard volume term derived in \cite{Henningson:1998gx,Henningson:1998ey}. Note that the second counterterm can also be written as
\be
 \frac{1}{16 \pi G_{d+1}} \int d^{d} x \sqrt{-\gamma} \a_{1/2} \frac{1}{(d-1)^2} \sum_{I}^{d-1}\phi_I \nabla_{i} \left ( \frac{\partial^i \phi_I}{ \sqrt{ (\partial \phi_I)^2}} \right ),
\ee
using partial integration.

Let us now restrict to $d=2$ and calculate the conformal anomaly and the renormalised mass. We need only consider the following additional terms in the metric
\be
g_{ij} = g_{(0) ij} + \rho^{1/2} g_{(1) ij} + \rho (g_{(2) ij} + {\rm ln} (\rho) h_{(2) ij} ) + \cdots
\ee
The $(\rho \rho)$ Einstein equation at order $1/\rho$ fixes ${\rm Tr} (g_{(0)}^{-1} h_{(2)}) = 0$. From \eqref{tr-ric} we obtain
\bea
&& {\cal R} + 2 (d-1) {\rm Tr} (g_{(0)}^{-1} g_2) - (d-1) {\rm Tr} (g_{(0)}^{-1} g_1)^2 - \frac{1}{4} ({\rm Tr}(g_{(0)}^{-1} g_1))^2 \\ 
&& = -  a_{1/2} \sum_{I} \frac{(2d-1)}{4 (d-1) Y_{(0)}} g_{(1)}^{jk} \partial_{j} \phi_{(0)I} \partial_k \phi_{(0) I}, \nn
\eea
where the indices of $g^{ij}_{(1)}$ have been raised with $g_{(0)}^{-1}$. This equation can be solved to give
\be
{\rm Tr} (g_{(0)}^{-1} g_{(2)}) = - \frac{\cal R}{2 (d-1)} + \frac{a_{1/2}^2}{8 (d-1)^3} (\sum_I Y_{(0) I})^2 + \frac{a_{1/2}^2 (2d-3) }{8(d-1)^3} \sum_{I,J} \frac{X_{(0) I}^{ij}  X_{(0) J ij}}{Y_{(0) I} Y_{(0) J}} 
\ee
with
\be
X_{(0) Iij} = \partial_{i} \phi_{(0)I} \partial_j \phi_{(0) I}.
\ee
The term involving the Ricci scalar agrees with \cite{deHaro:2000xn} \footnote{Note that their curvature conventions differ from ours.}. In $d=2$ there is only one species of scalar field and the expression simplifies to give 
\be
{\rm Tr} (g_{(0)}^{-1} g_{(2)}) = - \frac{\cal R}{2}  + \frac{1}{4} a_{1/2}^2 \partial^{i} \phi_{(0)} \partial_{i} \phi_{(0)}.
\ee
The divergence of $g_{(2)}$ is determined using the order one terms in the $(\rho i)$ Einstein equations 
\be
\nabla_{(0)}^{i} g_{(2) ij} = \frac{3 a_{1/2}}{2 Y_{(0)}} \partial_{j} \phi_{(0)} \phi_{(3)} + \cdots \label{diff}
\ee
In $d=2$ the rest of $g_{(2) ij}$ is not fixed, being related to the expectation value of the energy momentum tensor. The logarithmic term $h_{(2) ij}$ vanishes; one can show this using the $(ij)$ equations at order one. Solving the scalar field equation at order $\rho$ gives $\td{\phi}_{(3)} = 0$, i.e. the logarithmic term in the scalar field expansion vanishes. 

Using these expressions one can show that there is a logarithmic contribution to the onshell action in $d=2$
\be
S_{{\rm div}} = \frac{1}{16 \pi G_{3}} \int d^2 x \sqrt{-g_{(0)}} \ln \epsilon \left ( {\cal R} (g_{(0)}) \right )
\ee
which can be removed by the logarithmic counterterm
\be
S_{\rm ct} = - \frac{1}{16 \pi G_{3}} \int d^2 x \sqrt{-\gamma} \ln \epsilon \left ( {\cal R} (\gamma) \right ). \label{act3}
\ee
Note that the metric variation of this term is zero, in agreement with the fact that $h_{(2) ij} = 0$. 
 
The total action in $d=2$ is therefore the sum of \eqref{act1}, \eqref{act2} and \eqref{act3}:
\be
S_{\rm ren} = S_{\rm bare} + S_{\rm ct} + S_{\rm finite},
\ee
where the last term denotes finite counterterms, i.e. scheme dependent terms. The most relevant such term, which we will discuss further below is
\be
S_{\rm finite} = \frac{\gamma_s}{16 \pi G_3} \int d^2 x \sqrt{-\gamma} (\partial \phi)^2, \label{scheme}
\ee
where $\gamma_s$ is an arbitrary c-number. 

Varying the renormalised onshell action with respect to $g_{(0)ij}$ gives the renormalised stress energy tensor, defined as 
\be
\langle T_{ij} \rangle = \frac{2}{ \sqrt{\rm {det} (g_{(0)})}} \frac{\delta S_{\rm E ren}}{\delta g_{(0)}^{ij}} = {\rm Lim}_{\epsilon \rightarrow 0 } \left ( \frac{1}{\epsilon^{d/2 -1}} T_{ij} [ \gamma ] \right ). 
\ee
Here we have analytically continued to Euclidean signature, under which $i S \rightarrow - S_E$ with $S_E$ the Euclidean action. From the terms in the action involving only the metric and extrinsic curvature we obtain
\be
\langle T_{ij} \rangle =  \frac{1}{8 \pi G_3} \left (  g_{(2) ij} - {\rm Tr}(g_{(0)}^{-1} g_{(2)}) g_{(0) ij} + \frac{1}{2} {\rm Tr}(g_{(0)}^{-1} g_{(1)})^2 g_{(0) ij} 
- \frac{1}{2} {\rm Tr} (g_{(0)}^{-1} g_{(1)} ) g_{(1) ij}\right ),
\ee
in agreement with \cite{deHaro:2000xn} when $g_{(1)ij} = 0$, 
and the terms involving the scalar field give
\be
\langle T_{ij} \rangle = \frac{a_{1/2}^2}{8 \pi G_3} \left ( \frac{1}{4} \partial_k \phi_{(0)} \partial^k \phi_{(0)} g_{(0) ij} - \frac{1}{4} \partial_{i} \phi_{(0)} \partial_j \phi_{(0)} \right ).
\ee
Combining these gives
\be
\langle T_{ij} \rangle = \frac{1}{8 \pi G_{3}} \left ( g_{(2) ij} + \frac{\cal R}{2} g_{(0) ij} + a_{1/2}^2 \left ( - \frac{3}{4} \partial_{i} \phi_{(0)} \partial_j 
\phi_{(0)} + \frac{1}{2} (\partial \phi_{(0)})^2 g_{(0) ij}  \right ) \right ).
\ee
Note that the additional finite counterterm \eqref{scheme} contributes an additional traceless scheme dependent term
\be
\langle T^{s}_{ij} \rangle = \frac{\gamma_s}{8 \pi G_3} \left (  - \partial_{i} \phi_{(0)} \partial_j 
\phi_{(0)} + \frac{1}{2} (\partial \phi_{(0)})^2 g_{(0) ij}  \right ) 
\ee
The trace gives
\be
\langle T_{i}^{i} \rangle = \frac{\cal R}{16 \pi G_3},
\ee
i.e. the scalar field drops out of the conformal anomaly, as does the scheme dependent term. 

The operators dual to the scalar fields similarly have expectation values defined as
\be
\langle {\cal O}_I \rangle = \frac{1}{ \sqrt{ {\rm det} g_{(0)}}}  \frac{\delta S_{\rm ren}}{\delta \phi_{(0)I}} =  {\rm Lim}_{\epsilon \rightarrow 0} \left (  \frac{1}{\epsilon \sqrt{\gamma}} \frac{\delta S_{\rm ren}}{\delta \phi_{I}} \right ) . 
\ee
Again this is defined in Euclidean signature. Computing this quantity in $d=2$ gives
\be
\langle {\cal O} \rangle = - \frac{3 a_{1/2}}{16 \pi G_{3}}\frac{\phi_{(3) }}{Y_{(0)}} + \langle {\cal O}^s \rangle. \label{scalar_vev}
\ee
As anticipated, we note that that the expectation value is the normalizable mode, divided by $Y_{(0)}$. The total scaling weight is therefore two: the dual operator is marginal, despite the fact that the normalizable modes occur at order three in the Fefferman-Graham expansion. The term $\langle {\cal O}^s \rangle$ denotes scheme dependent contributions; for the specific counterterm \eqref{scheme} we obtain
\be
\langle {\cal O}^{s} \rangle = \frac{\gamma_s}{8 \pi G_3} \Box \phi_{(0)},
\ee
where $\Box$ is the d'Alambertian in the metric $g_{(0)}$. 

Using \eqref{diff} one can obtain the diffeomorphism Ward identity
\be
\nabla_{(0)}^{i} \langle T_{ij} \rangle = \partial_j \phi_{(0)} \langle {\cal O} \rangle. 
\ee
Switching on a source for $\phi_{(0)}$ which depends on the spatial coordinates, but not the time coordinate, allows momentum to be dissipated while preserving energy conservation. 

For $d > 2$ we would need to work out the series expansion to higher order and compute additional counterterms. However, the general structure will be analogous: 
\bea
\langle T_{ij} \rangle &=& \frac{d}{16 \pi G_{d+1}} g_{(d)ij} + \cdots \label{hol:d} \\
\langle {\cal O}_{I} \rangle &=& - \frac{d+1}{16 \pi G_{d+1}} \frac{a_{1/2} \phi_{(d+1)I}}{Y_{(0)I}} + \cdots, \nn
\eea
with the ellipses denoting terms local in  $g_{(0)ij}$ and $\phi_{(0) I}$. The leading term in the stress tensor involving the normalizable term in the metric expansion is as in \cite{deHaro:2000xn}.

\subsection{Thermodynamics of brane solutions} \label{sec:firstlaw}

The analysis above allows us to evaluate the onshell action on black brane solutions \eqref{bhsol} in $d=2$. In three bulk dimensions the metric  \eqref{bhsol} can be rewritten in Fefferman-Graham coordinates as
\be
ds^2 = \frac{d\rho^2}{4 \rho^2} + \frac{1}{\rho} \left (1 + \frac{\alpha \rho^{1/2}}{4} \right )^4 \left  (  -dt^2 \left (1 - \frac{\a \rho^{1/2}}{(1 + \frac{\a \rho^{1/2}}{4})^2}  \right ) + dx^2 \right ). \label{fg-form}
\ee
where we introduce the shorthand notation $\a = a_{1/2} c_{1/2}$. The general solution with parameter $m_0 \neq 0$ is
\be
ds^2 = \frac{1}{z^2} \left ( - ( 1 - \alpha z - m_0 z^2) dt^2 + \frac{dz^2}{(1 - \alpha z - m_0 z^2 )} + dx^2 \right )
\ee
and it can be rewritten in Fefferman-Graham coordinates using the transformation
\be
\rho^{\frac{1}{2}} = \frac{z}{1 - \frac{1}{2} \alpha z + \sqrt{1 - \alpha z - m_0 z^2}}.
\ee
The horizons of the general solution are located at
\be
z_{\pm} = - \frac{\alpha}{2 m_0} \pm \frac{1}{2 m_0} \sqrt{ \alpha^2 + 4 m_0}.
\ee
We noted previously that when $m_0 = 0$ a horizon exists only for $\alpha > 0$. When we allow for $m_0 \neq 0$, horizons exist provided that 
\be
m_{0} \ge - \frac{\alpha^2}{4}; \qquad \alpha \ge 0. 
\ee
When $m_0 = - \frac{\alpha^2}{4}$ the black brane is extremal. The entropy and temperature of the black brane are given by
\be
{\cal S} = \frac{V_1}{4 G_3 z_{+}} \qquad T = \frac{m_0 (z_{+} - z_{-})}{4 \pi}.
\ee
We can compute the free energy from the renormalised onshell action. We need to distinguish between the case in which the metric
has a horizon and the case in which it has a singularity as $z \rightarrow \infty$. 

{\bf No horizon:} In the latter case there are no contributions to the onshell action from the $z \rightarrow \infty$ limit of the volume integral and the renormalised Euclidean onshell action is
\be
S_{E}^{\rm onshell} = - \frac{\beta_T  V_1}{16 \pi G_3} \left ( m_0 + \frac{\alpha^2}{4} + \gamma_s c_1^2 \right ), 
\ee
where we have included the finite counterterm \eqref{scheme}. Here $\beta_T$ is the (arbitrary) period of the imaginary time direction.

Computed on \eqref{bhsol} the expectation value of the scalar operator is zero, which implies that the stress energy tensor is conserved. 
Using \eqref{fg-form} we can read off
\be
g_{(2) tt} = \frac{1}{8} \alpha^2 + \frac{1}{2} m_0; \qquad
g_{(2) xx} = \frac{3}{8} \alpha^2 + \frac{1}{2} m_0,
\ee
and therefore the conserved mass is 
\be
{\cal M} = \int dx \langle T_{tt} \rangle = \frac{V_1}{16 \pi G_3} \left (  (- \frac{3}{4} \a^2 - \gamma_s c_{1/2}^2 + m_0 \right ). \label{mass1}
\ee
The thermodynamic relation
\be
- S_{E}^{\rm onshell} = \beta_T F = \beta_T {\cal M},
\ee
where $F$ is the free energy, is satisfied provided that the coefficient of the scheme dependent term is
\be
\gamma_s = - \frac{1}{2} a_{1/2}^2.
\ee
Therefore the scheme dependence is fixed by imposing the thermodynamic relation. Note that these solutions are not however physical as the free energy is unbounded from below; they are analogous to negative mass Schwarzschild and indeed when we consider the limit $\alpha = 0$, $m_{0} < 0$ we recover negative mass BTZ.

{\bf Black brane:} In the black brane case the analysis is similar, but there are contributions to the onshell action from the horizon limit of the volume integral, resting in an onshell action
\be
S_{E}^{\rm onshell} = - \frac{\beta_T  V_1}{16 \pi G_3} \left ( m_0 + \frac{\alpha^2}{4} + \gamma_s c_{1/2}^2 + \left ( \frac{\alpha}{z_+} - \frac{2}{z_{+}^2} \right ) \right ), 
\ee
Here $V_{1}$ is the regulated length of the spatial direction and $\beta_T$ is the inverse temperature, which is no longer arbitrary. The conserved mass is as given in \eqref{mass1}. The thermodynamic relation 
\be
F = {\cal M} - T {\cal S}
\ee
is again satisfied provided that 
\be
\gamma_s = - \frac{1}{2} a_{1/2}^2. 
\ee
These solutions have a mass which is bounded from below
\be
{\cal M} \ge - \frac{V_1}{32 \pi G_3} \alpha^2,
\ee
with the bound being saturated by extremal black branes. This is most easily shown by rewriting the mass as
\be
{\cal M} = \frac{V_1}{16 \pi G_3} \left ( - \frac{1}{2} \alpha^2 + 4 \pi^2 T^2 \right ),
\ee
with the first term being a Casimir term and the second term showing the expected temperature dependence for a dual 2d conformal field theory. 
To derive the first law, note that one should vary the entropy and mass with respect to the temperature $T$, keeping the parameter $\alpha$ fixed. Then
\be
d {\cal M} = \frac{\pi V_1}{2 G_3} T d T.
\ee
In varying the entropy, it is useful to note that 
\be
{d z_{+}}_{| \alpha} = - 2 \pi z_{+}^2 dT
\ee
and therefore
\be
d {\cal S} = \frac{V_1}{4 G_3} \left ( \frac{-dz_{+}}{z_{+}^2} \right ) = \frac{\pi V_1}{2 G_3} dT,
\ee
which implies that the first law  $d {\cal M} = T d {\cal S}$ is indeed satisfied. 

\bigskip

The black brane solution in $d \ge 2$ with parameter $m_0 \neq 0$ is
\be
ds^2 = \frac{1}{z^2} \left ( - ( 1 - \frac{\alpha}{(d-1)} z - m_0 z^d) dt^2 + \frac{dz^2}{(1 - \frac{\alpha}{(d-1)} z - m_0 z^d )} + dx^2 \right ). \label{brane-d}
\ee
Let us denote the location of the outer horizon as $z_{0}$; it is the smallest value of $z$ at which the blackening function has a zero. The black brane has an extremal horizon when the blackening function has a double zero at $z_0$; this occurs when
\be
z_{0} = \frac{d}{\alpha}; \qquad  m_0 = - \frac{1}{(d-1) d^d} \alpha^d
\ee
and (at fixed $\alpha$) smaller values of $m_0$ give naked singularities. The entropy and temperature are given by
\be
{\cal S} = \frac{V_{d-1}}{4 G_{d+1} z_0^{d-1}}; \qquad T = \frac{1}{4 \pi} \left ( \frac{d}{z_0} - \alpha \right ).
\ee
Using \eqref{hol:d} the mass is given by
\be
{\cal M} = \frac{V_{d-1}}{16 \pi G_{d+1}} \left ( (d-1) m_0  + \lambda \alpha^{d} \right ),
\ee
where $\lambda$ is a constant which can only be determined by computing the local terms in \eqref{hol:d}. 
The first law is proved as follows: to vary ${\cal M}$ at fixed $\alpha$ we use
\be
d m_{0} = - \frac{dz_0}{z_0^d} \left ( \frac{d}{z_0} - \alpha \right ) = - 4 \pi T \frac{dz_0}{z_{0}^{d}}.
\ee
Therefore
\be
d {\cal M} = - \frac{V_{d-1}}{4 G_{d+1}}  \frac{ (d-1) T dz_0}{z_{0}^d} = T d {\cal S}. 
\ee

\subsection{Two point functions} \label{2pf}

Solvng \eqref{lin-eq} and \eqref{hol-ren} one can compute the two point function of the scalar operator in the conformal vacuum. The linearised equation 
\be
\bar{\nabla}^{\mu} \left ( \frac{1}{\sqrt{ (\partial \phi_I)^2}}  \bar{\nabla}_{\mu} \phi_I  \right ) = 0, \label{line-eq2}
\ee
admits solutions whose asymptotic expansion around the conformal boundary is given by \eqref{s1}. The two independent coefficients, $\phi_{(0)I}(x)$ and $\phi_{(d+1)I}(x)$, are related when one solves the equation throughout the bulk, imposing regularity everywhere. Regularity is however made more subtle by the fact that the backreaction on the metric occurs at the same order. The two point function for the dual scalar operator is then given by
\be
\langle {\cal O}_{I} (x) {\cal O}_I (y) \rangle =  {\rm Lim}_{\phi_{(0)I} \rightarrow 0} \left ( \frac{ (d+1) a_{1/2}}{16 \pi G_{d+1}} \frac{\delta \left ( \phi_{(d+1)I} (x)/Y_{(0)I} (x) \right ) }{\delta \phi_{(0) I} (y)}  + \cdots \right )
\ee
where the ellipses denote contact terms. We also need to take into account the fact that the backreaction on the metric is at linear order in the amplitude of the scalar field and therefore
\be
\langle T_{ij} (x) {\cal O}_I (y) \rangle = - {\rm Lim}_{\phi_{(0)I} \rightarrow 0} \left ( \frac{d}{16 \pi G_{d+1}} 
\frac{\delta g_{(d)ij}(x)}{\delta \phi_{(0)I} (y)} + \cdots \right ) \label{to2pf}
\ee 
does not automatically vanish. (Again the ellipses denote contact terms.)

Equation \eqref{line-eq2} is hard to solve. Since it is a non-linear equation, one cannot Fourier transform along the $x^i$ directions. One can solve for a single Fourier mode, i.e. letting
\be
\phi_I(z,x^i) = \td{\phi}_I(z, k_i) e^{i k^i x_i}
\ee
the equation becomes
\be
z^{d+1} \partial_z \left ( \frac{1}{z^d} \frac{\partial_z \td{\phi}_I}{\sqrt{ (\partial_z \td{\phi}_I)^2 + k_i k^i \td{\phi}_I^2}} \right ) = 0,
\ee
where we work with the usual Poincar\'{e} coordinates for $AdS_{d+1}$, namely
\be
ds^2 = \frac{1}{z^2} \left ( dz^2 + dx^i dx_i \right ).
\ee
The equation can then immediately be integrated once to give
\be
\frac{\partial_z \td{\phi}_I}{z^d} = \lambda k^d \sqrt{ (\partial_z \td{\phi}_I)^2 + k^2 \td{\phi}_I^2}
\ee
where $k^2 = k_i k^i$ and $\lambda$ is a dimensionless constant. This equation can be integrated to give
\be
\td{\phi}_I(z,k_i) = \td{\phi} (k_i) e^{i k^i x_i} {\exp} \left ( \lambda \int \frac{k^{d+1} z^d dz}{(1 - \lambda^2 (kz)^{2d})^{1/2}} \right ).
\ee
One can fix the constant $\lambda$ by imposing regularity and thereby relate the non-normalizable and normalizable modes in the asymptotic expansions. 
However, since one cannot linearly superpose such Fourier modes, one cannot use these solutions to compute the two point function.

Let us now consider perturbations about a background solution with non-vanishing scalar fields. The linearised problem is perfectly well-defined when the square root terms are expanded around any non-vanishing background. However, the metric and scalar field fluctuations are coupled and the equations of motion need to be diagonalised. f the scalar field fluctuations were decoupled from those of the metric, $a_{1/2}$ would need to be positive for the fluctuations to have the correct sign kinetic term (and hence, correspondingly, positive norm correlation functions in the holographically dual theory). Since the metric and scale fluctuations are coupled one cannot immediately conclude that the sign of the coefficient $a_{1/2}$ in \eqref{pot} must be positive.

We can compute the linearised equations of motion around \eqref{brane-d} as follows. We perturb the metric as $g_{\mu \nu} \rightarrow g_{\mu \nu} + h_{\mu \nu}$ and the scalar fields as $\phi_{I} \rightarrow \phi_{I} + \delta \phi_{I}$. Then the linearized scalar field equation is 
\begin{equation}
	\nabla_{\mu} \left(\frac{\nabla^{\mu} \delta\phi_I}{\sqrt{(\del\phi_I)^2}}\right) - \frac{1}{2}\frac{(\nabla^{\mu}\phi_I)}{\sqrt{(\del\phi_I)^2}}\nabla_{\mu} \left(\frac{\delta(\del\phi_I)^2}{(\del\phi_I)^2}\right) = \nabla_{\mu}\left( \frac{1}{\sqrt{(\del\phi_I)^2}}(h^{\mu \nu} - \frac{1}{2}h g^{\mu \nu})\nabla_{\nu} \phi_I\right)
	\label{eqn:phi_perturbed_eom}
\end{equation}
where we define $h = g^{\mu \nu} h_{\mu \nu}$. 
 
The linearised  Einstein equations are
\begin{equation}
	\delta R_{\mu \nu} = - d h_{\mu \nu} + \delta\bar{T}^{(1/2)}_{\mu \nu}
\end{equation}
where 
\begin{equation}
	\delta R_{\mu \nu} = -\frac{1}{2}	\Box h_{\mu \nu} - \frac{1}{2}\nabla_{\mu} \nabla_{\nu} h 
	+ \frac{1}{2}\nabla^{\rho} \nabla_{\mu} h_{\rho \nu} + \frac{1}{2}\nabla^{\rho} \nabla_{\nu} h_{\rho \mu}
\end{equation}
and 
\begin{align}
		\delta \bar{T}^{(1/2)}_{\mu \nu} &=  \frac{a_{1/2}}{2}\sum_{I=1}^{d-1}\frac{1}{\sqrt{(\del \phi_I)^2}}\left[
		\del_{\mu} \phi_I \del_{\nu} \delta \phi_I + \del_{\mu} \delta \phi_I \del_{\nu} \phi_I + \frac{1}{d-1}(\delta(\del\phi_I)^2g_{\mu \nu} + (\del\phi_I)^2 h_{\mu \nu})\right. \nonumber \\
		& \qquad \qquad \qquad\qquad\qquad \left.-\frac{1}{2}\frac{\delta (\del\phi_I)^2}{(\del\phi_I)^2}\left(\del_{\mu} \phi_I \del_{\nu} \phi_I + \frac{1}{d-1}(\del\phi_I)^2g_{\mu \nu}\right)
	\right]
\end{align}
where $\delta (\del\phi_I)^2 = 2 g^{\mu \nu}\del_\mu\phi_I \del_\nu\delta\phi_I - h^{\mu \nu}\del_\mu \phi_I \del_\nu \phi_I$.
These equations are complicated, since the scalar field profiles break relativistic invariance in the $d$ directions $(t,x^I)$.
One can however show that it is consistent to switch on only 
$h_{tI}(t,z)$, $h_{zI}(t,z)$ and $\delta \phi_I (t,z)$ for a given value of $I$; such perturbations suffice to compute the autocorrelation function. The three non-trivial equations are then the $(tI)$ Einstein equation, the $(zI)$ Einstein equation and the scalar field equation. Furthermore, one can choose a gauge $h_{zI} = 0$, and show explicitly that the two Einstein equations are compatible with each other. The resulting two equations are then simply
\bea
&& \delta \phi_I'' + \left ( \frac{F'}{F} - \frac{d}{z} \right ) \delta \phi_I' - \frac{1}{F^2} \partial_t^2 \delta \phi_I = \frac{ c_{1/2}}{F^2} \partial_t H_{tI}; \label{eq11} \\
&& \frac{1}{2 F} \partial_t H_{tI}' - \frac{a_{1/2}}{2z} \delta \phi_I' = 0,  \nn
\eea
where we have defined $H_{tI} = h_{tI} z^2$. 
One can eliminate $H_{tI}$ by taking the $z$ derivative of the first equation and using the second equation. 
Defining
\be
\z_I = z^{-d} F \delta \phi_I'
\ee
the resulting equation is
\be
z^{-d} \left ( z^d F \z_I' \right )' - \frac{1}{F} \partial_t^2 \z_I = \frac{1}{z} a_{1/2} c_{1/2} \z_I. \label{z1}
\ee
Note that the term on the right hand side should not be interpreted as a mass term in the usual sense, as it does not control the powers in the  asymptotic expansion of the field $\z_I$ as $z \rightarrow 0$. Indeed $\z_I$ admits two independent solutions
\be
\z_I = \z_{I (1-d)} (t) z^{1- d} + \cdots + \z_{(0)I} (t) + \frac{1}{d} c_{1/2} a_{1/2} \z_{(0)I} (t)  z + \cdots
\ee
which in turn correspond to an expansion 
\be \label{relp}
\delta \phi_I = \delta \phi_{(0)I} (t) + \cdots  + \delta \phi_{(d+1)I} (t) z^{d+1}+  \cdots,
\ee
in agreement with the full non-linear expression given in \eqref{s1}. 
We can give insight into the powers arising in this expansion as follows. For a massive scalar field the onshell action can be expressed as 
\be
- \int d^{d+1} x \sqrt{-g} \left ( \nabla^{\mu} \phi \nabla_{\mu} \phi + m^2 \phi^2 \right ) = - \int d\Sigma^{\mu} \phi \nabla_{\mu} \phi \approx - \int d^{d} x \frac{1}{z^{d-1}} \phi \partial_z \phi 
\ee
where we use the fact that the metric is asymptotically anti-de Sitter. Scale invariance requires that the non-normalizable mode of $\phi$ scales as $z^{d-\Delta}$ and acts as the source for an operator of dimension $\Delta$, and the field equation determines that $\Delta (\Delta - d) = m^2$; the normalizable mode of $\phi$ is related to the expectation value of this operator and scales as $z^{\Delta}$ \cite{Gubser:1998bc,Witten:1998qj}. 

Now let us turn to the square root model. Although the scalar field fluctuation is coupled to the metric fluctuation, the latter only affects subleading terms in the scalar field expansion near the conformal boundary: the leading asymptotics are controlled by the first two terms of the first equation in \eqref{eq11}. Therefore we can consider only the scalar field part of the action, i.e. 
\be
- a_{1/2} \int d^{d+1} x \sqrt{-g} \sqrt{ (\partial \phi_I + \partial \delta \phi_I)^2}
\ee 
Since this action is expanded around a solution of the equations of motion, the linear term in $\delta \phi_I$ automatically vanishes onshell. The onshell action can be expressed as a boundary term
\be
- a_{1/2} \int d\Sigma^{\mu} \left ( \frac{\delta \phi_I \partial_{\mu} \delta {\phi_1}}{2 \sqrt{ (\partial \phi_I)^2}} \right )
\ee
which gives
\be
- a_{1/2} \int d^{d} x \frac{1}{c_{1/2} z^{d}} \delta \phi_I \partial_z \delta \phi_I. 
\ee
Suppose the non-normalizable mode of $\delta \phi_I$ scales as $z^{0}$ and the normalizable mode scales as $z^{\Delta}$. The boundary term gives a contribution of order $z^0$ when $\Delta = d + 1$, in agreement with \eqref{relp}. 

The two point autocorrelation function can now be computed by solving \eqref{z1} for an arbitrary boundary source $\z_{I(1-d)}(t)$ subject to regularity; this will determine the subleading coefficients in \eqref{relp} and then
\be
\langle {\cal O}_{I}(t) {\cal O}_I (t') \rangle =  \frac{(d+1) a_{1/2}}{16 \pi c_{1/2} G_{d+1}} \frac{\delta \phi_{(d+1)I}(t)}{\delta \phi_{(0)I}(t')}.
\ee
It is straightforward to solve \eqref{eq11} in the limit $m_{0} = 0$ with $c_{1/2} \rightarrow 0$, as the fields decouple and therefore one can immediately solve for the scalar field. In the frequency domain we obtain
\be
\delta \phi_{I} (\omega) = \delta \phi_{(0)I} (\omega) (\omega z)^{(d+1)/2} K_{(d+1)/2} (\omega z),
\ee
where the Bessel function is normalised so that the asymptotic expansion takes the form \eqref{relp} and implicitly we are now working in Euclidean signature. Working in $d=2$, where the complete expression for the one point function was calculated in \eqref{scalar_vev}, we obtain
\be
\langle {\cal O}_{I}(\omega,0 ) {\cal O}_I (- \omega,0) \rangle =  \frac{a_{1/2}}{16 \pi G_{3}} \frac{\omega^3}{c_{1/2}} + {\mathcal O}(c_{1/2}^0),
\ee
where we work in mixed representation, i.e. frequency space and position space for the spatial coordinate. 
This expression is not analytic as $c_{1/2} \rightarrow 0$, i.e. as the background profile for the scalar field is switched off. 
Recall that the general expression for the Fourier transform of a polynomial in $d$ dimensions is
\be
\int d^{d}x e^{- i \vec{k} \cdot \vec{x}} ( | x|^2)^{-\lambda} = \pi^{d/2} 2^{d -2 \lambda} \frac{\Gamma(d/2 - \lambda)}{\Gamma(\lambda)} (|k|^2)^{\lambda - d/2},
\ee
which is valid when $\lambda \neq (d/2 + n)$, where $n$ is zero or a positive integer. Transforming back to the (Euclidean) time domain gives
\be
\langle {\cal O}_{I}(t) {\cal O}_I (t') \rangle = \frac{3 a_{1/2}}{64 \pi^2 G_3} \frac{1}{ c_{1/2} |t|^4} + {\mathcal O}(c_{1/2}^0).
\ee
This condition is consistent with a positive norm provided that $a_{1/2} c_{1/2} > 0$. The off-diagonal correlation function \eqref{to2pf} is of order $c_{1/2}$ or smaller.

\section{Phenomenological models} \label{sec:four}

In this section we will explore the properties of the following model:
\begin{equation}
	S = \frac{1}{16 \pi G_{d+1}}  \int \dd^{d+1}x \sqrt{-g}\left(R + d(d-1) - \frac{1}{4}F^2 - \sum_{I=1}^{d-1}(a_{1/2}\sqrt{(\del \psi_I)^2} + a_{1}(\del\chi_I)^2)\right)
\end{equation}
where for clarity we now label the two families of scalar fields as $\psi_I$ and $\chi_I$. The Einstein equation is
\begin{align}
	R_{\mu \nu} &= -d g_{\mu \nu} + \bar{T}_{\mu \nu} \\
		   &= -d g_{\mu \nu} +  \frac{1}{2}\left(F_{\mu \rho}{F_\nu}^\rho - \frac{1}{2(d-1)}F^2 g_{\mu \nu}\right) 
		   + \sum_{I=1}^{d-1} a_1 \del_\mu  \chi_I \del_\nu \chi_I \nonumber
		   \nonumber \\
		   & \quad + \sum_{I=1}^{d-1} \frac{a_{1/2}}{2\sqrt{(\del\psi_I)^2}}\left( \del_\mu \psi_I \del_\nu \psi_I + \frac{1}{d-1}(\del\psi_I)^2 g_{\mu \nu}\right) \nonumber 
\end{align}
and the other equations of motion are
\bea
\nabla^\mu \left(\frac{1}{\sqrt{(\del\psi_I)^2}} \nabla_\mu \psi_I\right) &=& 0; \\
\nabla_\mu \nabla^\mu \chi_I &=& 0; \qquad
\nabla_{\mu} F^{\mu \nu} = 0. \nn
\eea
The homogeneous and isotropic black brane solutions are given by
\be
ds^2 = \frac{1}{z^2} \left ( - F(z) dt^2 + \frac{dz^2}{F(z)} + dx \cdot dx \right ), \label{hibb}
\ee
with
\be
F(z) = 1 - m_0 z^d + \frac{\mu^2}{\gamma^2 z_0^{2(d-2)}} z^{2(d-1)} - \frac{1}{d-1}a_{1/2}c_{1/2}z - \frac{1}{d-2}a_1 (c_1 z)^2
 \label{eq:blackening_function}
\ee
where $\gamma^2 = 2(d-1)/(d-2)$ and $m_0$ is fixed by demanding that $F(z_0) = 0$. The Maxwell potential is
\be
A = \mu\left(1 - \frac{z^{d-2}}{z_0^{d-2}}\right) \mathrm{d}t 
\ee
and the scalar fields are given by
\be
\chi_{I} = c_{1} x^{I}; \qquad
\psi_I = c_{1/2} x^I. \label{scl:prof}
\ee
The temperature is given by 
\begin{equation}
	T = - \frac{F'(z_0)}{4\pi} = \frac{1}{4\pi}\left( \frac{d}{z_0} - \frac{(d-2)^2\mu^2 z_0}{2(d-1)} - a_1 c_1^2 z_0 - a_{1/2}c_{1/2} \right). \label{eq:temperature}
\end{equation}
The entropy is
\be
{\cal S} = \frac{V_{d-1}}{4 G_{d+1} z_{0}^{d-1}}
\ee
and the potential $\Phi$ and charge $Q$ are respectively
\be
\Phi = \mu; \qquad Q = \frac{(d-2) V_{d-1} \mu }{16 \pi G_{d+1} z_{0}^{d-2}}. \label{charge}
\ee
The charge density $q$ is given by $Q = q V_{d-1}$. 
The mass is given by 
\be
{\cal M} = \frac{(d-1)}{16 \pi G_{d+1}} \left ( m_0  + \cdots \right ) \label{Bb:mass}
\ee
where the ellipses denote terms involving the non-normalizable modes, i.e $\alpha \equiv a_{1/2} c_{1/2}$ and $\beta \equiv a_1 c_1^2$. This form for the mass is consistent with the mass for the standard Einstein-Maxwell system; one can show that the first law $d {\cal M} = T d {\cal S} + \Phi d Q$ is satisfied using analogous steps to those in section \ref{sec:firstlaw}. 
 
Systematic holographic renormalisation would be required to determine the terms in ellipses in the \eqref{Bb:mass} and the free energy. The scalar field profiles \eqref{scl:prof} are non-normalizable modes, associated with deformations of the dual field theory, and therefore the thermodynamically preferred state is that with lowest free energy at fixed $(c_{1/2},c_1)$. It is possible that the homogeneous black branes \eqref{hibb} are not the thermodynamically preferred state, particularly at low temperatures, but we will not investigate phase transitions here. Note that the near horizon geometry remains $AdS_2 \times R^{d-1}$, as in Reissner-Nordstr\"{o}m, and the entropy does not vanish at zero temperature. 

\subsection{Linearized perturbations}

We now consider linearised perturbations of the fields around the black brane backgrounds, such that
\begin{eqnarray}
	g_{\mu} &\to &g^{(0)}_{\mu \nu} + h_{\mu \nu}; \\
	A_\mu &\to&  A_\mu + \delta A_\mu; \nn \\
	\psi_I &\to & \psi_I + \delta \psi_I;  \nn \\
	\chi_I &\to & \chi_I + \delta \chi_I. \nn
\end{eqnarray}
In the following, all indices will be raised and lowered using the background metric $g_{\mu \nu}^{(0)}$ and its inverse unless otherwise stated, and all covariant derivatives $\nabla_\mu$ will be taken with respect to the background metric $g_{\mu \nu}^{(0)}$. Note that $g_{\mu \nu}^{(0)}$ in this section refers to the background black brane metric, and should not be confused with the leading term in the asymptotic Fefferman-Graham metric expansion, $g_{(0) ij}$. 

We consider homogeneous fluctuations of the following form
\bea
	h_{\mu\nu} &=& e^{-i \omega t} \frac{1}{z^2} H_{\mu\nu}(z); \qquad
	\delta A_\mu = e^{-i \omega t} a_\mu(z); \label{fluc:mode} \\
	\delta \psi_I &=& e^{-i \omega t} \Psi_I(z); \qquad
	\delta \chi_I = e^{-i \omega t} \mathcal{X}_I(z). \nonumber
\eea
It is straightforward to show that the perturbations $(h_{zI}, h_{tI},  \delta A_{I}, \delta \chi_{I}, \delta \psi_I)$ decouple. One can choose a gauge in which $h_{zI} = 0$, resulting in the following equations of motion. The scalar field equations are 
\begin{align}
	\mathcal{X}_I'' + \left[ \frac{F'}{F} - \frac{d-1}{z}\right] \mathcal{X}_I' + \frac{\omega^2}{F^2}\mathcal{X}_I - \frac{i \omega c_1}{F^2}H_{tI} &= 0 \label{eq:mm_chi} \\
	\Psi_I'' + \left[ \frac{F'}{F} - \frac{d}{z}\right] \Psi_I' + \frac{\omega^2}{F^2}\Psi_I - \frac{i \omega c_{1/2}}{F^2} H_{tI} &= 0 \label{eq:mm_psi} 
\end{align}
The $I$ component of the Maxwell equations and the $(zI)$ Einstein equations are
\begin{align}
a_I'' + \left[ \frac{F'}{F} - \frac{d-3}{z}\right]a_I' + \frac{\omega^2}{F^2}a_I - \frac{\mu(d-2)z^{d-3}}{z_0^{d-2}F} H_{tI}' &= 0 \label{eq:mm_a} \\
\frac{i \omega}{2F}H_{tI}' - \frac{a_{1/2}}{2z}\Psi_I' - a_1c_1 \mathcal{X}_I' - \frac{i \omega \mu(d-2) z^{d-1}}{2 F z_0^{d-2}} a_I &= 0. \label{eq:mm_einsten}
\end{align}
For each value of $I$, i.e. each spatial direction, we have four equations of motion. A homogeneous Maxwell 
field in the $I$th direction is coupled to both the metric perturbation $h_{tI}$ and perturbations of the scalar fields associated with this direction. 

It is convenient to eliminate $H_{tI}$ by taking the $z$-derivative of Equations \eqref{eq:mm_chi} and \eqref{eq:mm_psi}, and using Equation \eqref{eq:mm_einsten} to eliminate $H_{tI}'$ from the resulting equations. In the process of doing so it is also useful to introduce the following:
\begin{equation}
	\xi_I = \omega^{-1} z^{-(d-1)}F \mathcal{X}_I', \qquad \zeta_I = \omega^{-1} z^{-d} F \Psi_I'
\end{equation}
we can rewrite the remaining three field equations as:
\begin{eqnarray}
	z^{d-3}(z^{-(d-3)}F a_I')' + \frac{\omega^2}{F}a_I &=& (d-2)^2\mu^2\frac{z^{2(d-2)}}{z_0^{2(d-2)}}a_I - i(d-2)\mu a_{1/2} \frac{z^{2(d-2)}}{z_0^{d-2}} \zeta_I \nonumber \\
	&& \qquad \qquad - 2i(d-2) \mu a_1 c_1 \frac{z^{2(d-2)}}{z_0^{d-2}} \xi_I \label{eq:temperature_a_eom} \\
	z^{-d}(z^d F \zeta_I')' + \frac{\omega^2}{F}\zeta_I &=&  \frac{1}{z} \frac{i(d-2)\mu c_{1/2}}{z_0^{d-2}}a_I + \frac{1}{z} a_{1/2}c_{1/2} \zeta_I + \frac{2}{z}a_1 c_1 c_{1/2} \xi_I \label{eq:temperature_zeta_eom} \\
	z^{-(d-1)}(z^{d-1} F \xi_I')' + \frac{\omega^2}{F} \xi_I &=& \frac{i(d-2) \mu c_1}{z_0^{d-2}}a_I + a_{1/2}c_1 \zeta_I + 2a_1 c_1^2 \xi_I \label{eq:temperature_xi_eom}
\end{eqnarray}

To analyse these equations further it is convenient to rewrite the perturbation field equations in a dimensionless form by making the coordinate change $r = \mu z$, and rescaling the sets of perturbations $\bar{a}_I = \mu^{d-2}a_I$, $\bar{\zeta}_I = \zeta_I / c_{1/2}$, $\bar{\xi}_I = \mu \xi_I / c_1$. After making these changes the field equations are simply:
\begin{eqnarray}
	\ddot{\bar{a}}_I + \left[\frac{\dot{F}}{F} - \frac{d-3}{r}\right] \dot{\bar{a}}_I + \frac{\bar{\omega}^2}{F^2}\bar{a}_I &=& \frac{1}{F}\left[ (d-2)^2 \left(\frac{r}{r_0}\right)^{2(d-2)} \bar{a}_I - i(d-2)\frac{r^{2(d-2)}}{r_0^{d-2}} \td{\a} \bar{\zeta}_I  \right . \nn \\
	&& \qquad \left . - i(d-2)\frac{r^{2(d-2)}}{r_0^{d-2}} \td{\b} \bar{\xi}_I \right] \\
	\ddot{\bar{\zeta}}_I + \left[\frac{\dot{F}}{F} + \frac{d}{r}\right] \dot{\bar{\zeta}}_I + \frac{\bar{\omega}^2}{F^2}\bar{\zeta}_I &=& \frac{1}{r F}\left[ \frac{i(d-2)}{r_0^{d-2}} \bar{a}_I + \td{\a} \bar{\zeta}_I + \td{\b} \bar{\xi}_I \right] \nn \\
		\ddot{\bar{\xi}}_I + \left[\frac{\dot{F}}{F} + \frac{d-1}{r}\right] \dot{\bar{\xi}}_I + \frac{\bar{\omega}^2}{F^2}\bar{\xi}_I &=& \frac{1}{F}\left[ \frac{i(d-2)}{r_0^{d-2}} \bar{a}_I + \td{\a}\bar{\zeta}_I + \td{\b} \bar{\xi}_I \right] \nn
\end{eqnarray}
where $\bar{\omega} = \omega / \mu$ and $r_0 = \mu z_0$. We have also introduced the shorthand $\td{\alpha} = a_{1/2} c_{1/2}/\mu$ and $\td{\beta} = 2 a_1 c_1^2/\mu^2$. 

It is immediately clear that these equations imply
\begin{equation}
	r^{-d}(r^d F \dot{\bar{\zeta}}_I - r^{d-1} F \dot{\bar{\xi}}_I) + \frac{\bar{\omega}^2}{F}\left(\bar{\zeta}_I - \frac{1}{r}\bar{\xi}_I\right) = 0
\end{equation}
and hence there exists a quantity
\begin{equation}
	\bar{\kappa}_I = r^d F \left( \bar{\zeta}_I - \frac{1}{r}\bar{\xi}_I \right)
\end{equation}
that is radially conserved in the $\bar{\omega} \to 0$ limit. One can easily show that, in the near boundary limit, $\bar{\kappa}_I = O(r^{d+1})$ and hence vanishes on the conformal boundary for all $\bar{\omega}$. We know that $\bar{\kappa}_I$ is also radially conserved in the $\bar{\omega} \to 0$ limit and hence it must vanish everywhere in this limit. Consequentially we find that
\begin{equation}
	\dot{\bar{\zeta}}_I = \frac{1}{r} \dot{\bar{\xi}}_I
\end{equation}
in the $\bar{\omega} \to 0$ limit.

The equations can be diagonalised to make manifest the existence of two massless modes, and hence two conserved quantities in the zero frequency limit. The eigenvectors are
\begin{eqnarray}
	\bar{\lambda}_{1I} &=&  \frac{\td{\b}}{\bar{B}} \left[  \bar{a}_I - \frac{i  \td{\a} r_0^{d-2}}{d-2}\left(\bar{\zeta}_I - \frac{1}{r} \bar{\xi}_I \right) - \frac{i(d-2)r^{2(d-2)}}{r_0^{d-2}}  \bar{\xi}_I \right]; \nn \\ 
	\bar{\lambda}_{2I} &=&  \frac{\td{\a}}{\bar{B}} \left[  \frac{1}{r} \bar{a}_I + \frac{i \td{\beta} r_0^{d-2}}{d-2}\left(\bar{\zeta}_I - \frac{1}{r} \bar{\xi}_I \right) + \frac{i(d-2)r^{2(d-2)}}{r_0^{d-2}} \bar{\zeta}_I \right] \\	
		\bar{\lambda}_{3I} &=& \frac{1}{\td{\a} \td{\b}\bar{B}} \left[  \frac{(d-2)^2}{r_0^{2(d-2)}} \bar{a}_I - \frac{i (d-2)}{r_0^{d-2}} \left( \td{\a} \bar{\zeta}_I + \td{\b} \bar{\xi}_I \right) \right]. \nn
	\end{eqnarray} 
where the quantity $\bar{B}$ is defined as
\begin{equation}
\bar{B}(r) =  \frac{\td{\a}}{r} + \td{\b} + (d-2)^2\left(\frac{r}{r_0}\right)^{2(d-2)}.	
\end{equation}
It is clear to see that $\bar{\lambda}_{1I}$ and $\bar{\lambda}_{2I}$ are massless modes as their equations of motion are:
\begin{align}
	\td{\beta} r^{d-3} \dot{\bar{\Pi}}_{1I} + \frac{\bar{\omega}^2 \bar{B}}{F} \bar{\lambda}_{1I} &= \frac{i \td{\alpha} \td{\beta} r_0^{d-2}}{d-2}r^{-d}\dot{\bar{\kappa}}_I \\
	\td{\alpha} r^{d-4} \dot{\bar{\Pi}}_{1/2 I} + \frac{\bar{\omega}^2 \bar{B}}{F} \bar{\lambda}_{2I} &= - \frac{i \td{\alpha} \td{\beta
	} r_0^{d-2}}{d-2}r^{-d}\dot{\bar{\kappa}}_I
\end{align}
where the two momenta are given by 
\begin{align}
	\bar{\Pi}_{1I} &= r^{-(d-3)} F \left[ \dot{\bar{a}}_I + \frac{i(d-2)r^{2(d-2)}}{r_0^{d-2}} \dot{\bar{\xi}}_I \right] \\
	\bar{\Pi}_{1/2I} &= r^{-(d-3)} F \left[ \dot{\bar{a}}_I + \frac{i(d-2) r^{2d-3}}{r_0^{d-2}} \dot{\bar{\zeta}}_I \right]
\end{align}
and are radially conserved in the $\bar{\omega} \to 0$ limit. It is clear that $\bar{\Pi}_{1/2 I} - \bar{\Pi}_{1I} = \frac{i(d-2)}{r_0^{d-2}} \bar{\kappa}_I$ and hence not only are they conserved in the $\bar{\omega} \to 0$ limit but they are also equal throughout the bulk in this limit.

To progress further we need to work out the asymptotic expansions near the conformal boundary for the various fluctuations fields under consideration. The three sets of fields $a_I, \zeta_I, \xi_I$ have both homogeneous and inhomogeneous contributions. Since the field equations \eqref{eq:temperature_a_eom}-\eqref{eq:temperature_xi_eom} are second order linear ODEs we expect each field to have two homogeneous contributions: one corresponding to a normalizable mode, and one non-normalizable. Since we are primarily interested in computing the conductivity we will turn off the non-normalizable modes for the scalar fields, which correspond to perturbing the sources for the dual operators in the field theory. (Note that the background solution still has sources for these operators.)

We make the following ansatz for the asymptotic expansions of the solutions to the full inhomogeneous equations:
\begin{align}
	a_I &= \sum_{k=0}^\infty z^k a_{I (k)} + \tilde{a}_{I (d-2)} z^{d-2} \log z + \cdots \\
	\zeta_I &= \sum_{k=0}^\infty z^k \zeta_{I(k)} + \tilde{\zeta}_{I(d-1)} z^{d-1} \log z  + \cdots \qquad
	\xi_I = \sum_{k=0}^\infty z^k \xi_{I(k)} + \tilde{\xi}_{I(d-2)} z^{d-2} + \cdots, \nn
\end{align}
where the logarithmic terms are included at the orders at which normalisable modes appear. The ellipses denote further logarithmic terms which we will not need here. Analysis of the field equations results in the following. From the Maxwell field equation we find that as usual $a_{I(1)} = \ldots = a_{I(d-3)} = 0$ and hence the leading order terms in $a_I$ are
\begin{equation}
	a_I = a_{I(0)} + a_{I(d-2)} z^{d-2} + O(z^{d-1})
\end{equation}
where we can identify the coefficients $a_{I(0)}$ as the dual first order perturbation to the gauge potential 
and $a_{I(d-2)}$ is related to the expectation value of the dual current. 

For the scalar fields the leading order terms in the expansions are
\begin{align}
	\zeta_I &= \zeta_{I(0)} + \frac{1}{d}c_{1/2}\left( i(d-2) \mu a_{I(0)} + a_{1/2} \zeta_{I(0)} + 2a_1 c_1 \xi_{I (0)} \right) z + \ldots \\
	\xi_I &= \xi_{I(0)} + \frac{1}{2d} c_1 \left( i(d-2) \mu a_{I(0)} + a_{1/2} \zeta_{I(0)} + 2a_1 c_1 \xi_{I(0)} \right) z^2 + \ldots. \nn
\end{align}
The dimensionless fields therefore have the following asymptotic behaviours:
\begin{eqnarray}
	\bar{a}_I &= & \bar{a}_{I(0)} + \bar{a}_{I(d-2)} r^{d-2} + O(r^{d-1}) \qquad
	\bar{\zeta}_I = \bar{\zeta}_{I(0)} + O(r) \\
	\bar{\xi}_I &=& \bar{\xi}_{I(0)} + O(r^2) \qquad
	\frac{1}{\bar{B}} =  \frac{r}{\td{\a}} - \frac{\td{\beta} r^2}{\td{\alpha}^2} + O(r^3) \nn \\
	\bar{\lambda}_{1I} &=& \bar{a}_{I(0)} + \ldots \qquad
	\bar{\Pi}_I = (d-2) \bar{a}_{I(d-2)} + O(r). \nn	
\end{eqnarray}

Note that the near horizon expansions of the fields are
\begin{align}
	a_I &= (z-z_0)^{i \omega / F'(z_0)} [ a_I^H + O((z-z_0)) ] \\
	\zeta_I &= (z-z_0)^{i \omega / F'(z_0)} [ \zeta_I^H + O((z-z_0)) ]  \nn \\
	\xi_I &= (z-z_0)^{i \omega / F'(z_0)} [ \xi_I^H + O((z-z_0)) ], \nn
\end{align}
with $(a_{I}^{H},\zeta_I^H,\xi_I^H)$ constants, or in terms of the dimensionless fields
\begin{align}
	\bar{a}_I &= (r - r_0)^{i \omega / \dot{F}(r_0)}\left[ \bar{a}_I^H + O((r - r_0)) \right] \\
	\bar{\zeta}_I &= (r- r_0)^{i \omega / \dot{F}(r_0)}\left[ \bar{\zeta}_I^H + O((r - r_0)) \right] \nn \\
	\bar{\xi}_I &= (r - r_0)^{i \omega / \dot{F}(r_0)}\left[ \bar{\xi}_I^H + O((r - r_0)) \right]. \nn
\end{align}
In the zero frequency limit, $\bar{\kappa}_I$ is zero, as it must vanish at the conformal boundary and is conserved. This in turn implies that 
$\bar{\zeta}_I^H = \bar{\xi}_I^H/r_0$ in the zero frequency limit. 

Putting this information together, one can show that a combination of the two massless modes is asymptotic to the source for the gauge field, i.e. 
\be
\bar{\lambda}_I= \bar{\lambda}_{1I} + \bar{\lambda}_{2I} = a_{I(0)} + O(z^2) \label{lam-asym}
\ee
The equation of motion for this mode is given by
\begin{equation}
	r^{-(d-3)}\left[\td{\b}\dot{\bar{\Pi}}_{1I} + \frac{\td{\a}}{r} \dot{\bar{\Pi}}_{1/2 I} \right] + \frac{\bar{\omega}^2 \bar{B}}{F}\bar{\lambda}_I = 0. \label{pi-mode}
\end{equation}
It is also clear that $\bar{\Pi}_{1I}$ and $\bar{\Pi}_{1/2 I}$ have equivalent near-boundary behaviour:
\begin{equation}
	\bar{\Pi}_{1I} = (d-2)\bar{a}_{I(d-2)} + O(z) = \bar{\Pi}_{1/2 I}  \label{pi-asym}
\end{equation}
which is not surprising as they only differ by a multiple of $\bar{\kappa}_I$ which vanishes in the $z \to 0$ limit.

It is clear from equation \eqref{pi-mode} that $\dot{\bar{\Pi}}_{1I} / \bar{\lambda}_{I} \sim O(\bar{\omega}^2)$ and $\dot{\bar{\Pi}}_{1/2 I} / \bar{\lambda}_I \sim O(\bar{\omega}^2)$. Similarly we know that $\bar{\Pi}_{1I} / \bar{\lambda}_I \sim O(\bar{\omega})$ and $\bar{\Pi}_{1/2 I} / \bar{\lambda}_I \sim O(\bar{\omega})$: these conditions are satisfied at the horizon due to ingoing boundary conditions and are conserved throughout the bulk by the field equations. We will use these properties in deriving the DC conductivity below. 

Finally we note that the $\bar{\lambda}_{3I}$ field equation is given by
\begin{align}
	0 &= \bar{B} \ddot{\bar{\lambda}}_{3I} + r_0^{-2d}r^{-5}\left[ ((d-2)^2(3d-5)r_0^4 r^{2d} + r_0^{2d} r^3 ( \td{\a}(d-2) + \td{\b} r(d-1))) F + r_0^{2d}r^5 \dot{F} \bar{B} \right] \dot{\bar{\lambda}}_{3I} \nn \\
	& \qquad + r^{-6}(- \td{\a} r^3 + 2(d-2)^3 r^{2d}r_0^{-2(d-2)})(r \dot{F} + (d-2)F)\bar{\lambda}_{3I} 
		+ \frac{\bar{\omega}^2 \bar{B}}{F}\bar{\lambda}_{3I} - \bar{B}^2 \bar{\lambda}_{3I}	 \\
	& \qquad + \frac{r_0^{-2d}r^{-7} F r^d}{\bar{B}}(r_0^{2d} \td{\alpha} \td{\beta} r^4 + (d-2)^2 r^{2d}r_0^4((2d-3)^2 \td{\a} + (2d-4)^2 \td{\b} r))\bar{\lambda}_{3I}  \nonumber \\
	& \qquad + \frac{r_0^{-2d}r^{-(d+5)}}{ \td{\alpha} \td{\beta} \bar{B}}\left( r_0^{2d} r^4 \td{\alpha} \td{\beta}(\bar{\Pi}_{1I} - \bar{\Pi}_{1/2 I}) - (d-2)^2 r^{2d}r_0^4((2d-3)\td{\a} \bar{\Pi}_{1/2 I} + (2d-4)\td{\b} r \bar{\Pi}_{1I})\right) \nonumber
\end{align}
and, in terms of the eigenmodes, $\bar{\Pi}_{1I}$ and $\bar{\Pi}_{1/2I}$ are given by
\begin{align}
	\bar{\Pi}_{1I} &= r^{-(d-3)} F \left[ \dot{\bar{\lambda}}_{1I} + \dot{\bar{\lambda}}_{2I} + \td{\alpha} \td{\beta}(2d-4)r^{2d-5}\bar{\lambda}_{3I} + \frac{(d-2)^2}{\td{\b}}\left(\frac{r}{r_0}\right)^{2(d-2)}\dot{\bar{\lambda}}_{1I} \right] \\
	\bar{\Pi}_{1/2 I} &= r^{-(d-3)} F \left[ \dot{\bar{\lambda}}_{1I} + \dot{\bar{\lambda}}_{2I} + \td{\a} \td{\b}(2d-3)r^{2d-5}\bar{\lambda}_{3I} + \frac{(d-2)^2}{\td{\a}}\left(\frac{r}{r_0}\right)^{2(d-2)}r\dot{\bar{\lambda}}_{2I} \right].
\end{align}
We will use the structure of these equations to derive the DC conductivity below. 

The equations of motion of the massive modes $\bar{\lambda}_{3I}$ are schematically given by
\begin{equation}
	L_3 \bar{\lambda}_{3I} + p_3(r) \bar{\lambda}_{3I} + \bar{\omega}^2 q_3(r) \bar{\lambda}_{3I} \sim \bar{\Pi}_{1I}
\end{equation}
where $L_3$ is a linear differential operator, and $p_3(r)$, $q_3(r)$ are functions of  the radial coordinate $r$ with no frequency dependence. The massless modes only couple to $\bar{\lambda}_{3I}$ via $\bar{\Pi}_{1I}$ and the $\bar{\lambda}_{3I}$ equations of motion hence yield that $\bar{\lambda}_{3I} \sim \bar{\Pi}_{I}$. The conjugate momentum takes the form
\begin{equation}
	\bar{\Pi}_{1I} = P(r) \dot{\bar{\lambda}}_{I} + Q(r) \bar{\lambda}_{3I}
\end{equation}
where again $P(r)$ and $Q(r)$ are functions with no frequency dependence. From this we deduce that $\bar{\Pi}_{1I} \sim \dot{\bar{\lambda}}_{I}$ and hence, recalling that $\bar{\Pi}_{1I}/\bar{\lambda}_{I} \sim O (\bar{\omega})$,
we know that $\dot{\bar{\lambda}}_{I}/\bar{\lambda}_I \sim O(\bar{\omega})$ as $\bar{\omega} \rightarrow 0$. We will use this property below in deriving the DC conductivity.

\subsection{DC conductivity}

In this section we will compute the DC limit of the optical conductivity. Since the equations in different spatial directions decouple, and are identical, we now restrict to perturbations in one of the boundary spatial directions which we will label by $x$. The optical conductivity in this direction is defined as
\begin{equation}
	\sigma_x(\omega) = \frac{\langle J_x \rangle}{i \omega A_{x (0)}}
	\label{eq:optical_conductivity}
\end{equation}
where $A_{x(0)}$ is the source for the $x$ component of the boundary current and $\langle J_x \rangle$ is its expectation value. The DC conductivity is defined by
\begin{equation}
	\sigma_{DC} = \lim_{\omega \to 0}\sigma_x (\omega),
\end{equation}
and due to the symmetry of the background and of the equations of motion takes the same value along all spatial directions. 
Note that the source is given by
\be
A_{x (0)} = a_{x (0)} e^{- i \omega t}
\ee
and the expectation value of the current is given by
\be
\langle J_x \rangle = (d-2) a_{x (d-2)} e^{-i \omega t} + \cdots 
\ee
where for notational simplicity we set $16 \pi G_{d+1} = 1$ for the remainder of this section. 

The holographic optical conductivity can be expressed in terms of the dimensionless fields as
\begin{equation}
	\frac{\sigma_x(\bar{\omega})}{\mu^{d-3}} = (d-2) \frac{\bar{a}_{x(d-2)}}{i \bar{\omega} \bar{a}_{x (0)}}
\end{equation}
with the DC conductivity being the $\bar{\omega} \to 0$ limit of this expression. Using our knowledge of the asymptotic behaviour of the massless modes and conserved quantities we now define the auxiliary quantity
\begin{equation}
	{\sigma}_{DC}(r) = \mu^{d-3} \lim_{\bar{\omega} \to 0} \frac{ \bar{\Pi}_{1x} }{i \bar{\omega} \bar{\lambda}_{x}}. \label{aux_dc} 
\end{equation}
From \eqref{lam-asym} and \eqref{pi-asym} it is clear that this quantity coincides  with the DC conductivity at the conformal boundary. However, we will now show that this function is conserved to leading order in $\omega$ and it can thus be evaluated at any value of the radius. 

Our equations of motion have a similar structure to those in \cite{Andrade2013,Blake2013} and therefore the proof that \eqref{aux_dc} is a conserved quantity closely follows their proofs. Radial conservation of \eqref{aux_dc} at leading order in the frequency requires that
\be
\frac{d}{dr} \left (  \frac{ \bar{\Pi}_{1x} }{ \bar{\lambda}_{x}} \right ) =\left (   \frac{ \dot{\bar{\Pi}}_{1x} }{ \bar{\lambda}_{x}}  - \frac{ \bar{\Pi}_{1x} }{ {\bar{\lambda}}_{x}} \frac{ \dot{\bar{\lambda}}_x }{ \bar{\lambda}_{x}} \right ) = O(\bar{\omega}^2). 
\ee
This result follows if the following three results hold: $\dot{\bar{\Pi}}_{1x}/ \bar{\lambda}_{x} \sim O(\bar{\omega}^2)$; 
${\bar{\Pi}}_{1x}/ \bar{\lambda}_{x} \sim O(\bar{\omega})$ and  $\dot{\bar{\lambda}}_{x}/ \bar{\lambda}_{x} \sim O(\bar{\omega})$ as $\bar{\omega} \rightarrow 0$. However, we already showed that all three conditions hold in the previous section. 

Since \eqref{aux_dc} is radially conserved we can calculate its value on the horizon giving
\begin{equation}
	\frac{\sigma_{DC}}{\mu^{d-3}} = r_0^{-(d-3)}\left(1 + \frac{(d-2)^2}{\td{\beta} + \td{\alpha} r_0^{-1}}\right).
\end{equation}
Reinstating all parameters explicitly we obtain
\begin{equation}
	\sigma_{DC} = z_0^{-(d-3)}\left( 1 + \frac{(d-2)^2 \mu^2}{2a_1 c_1^2 + z_0^{-1} a_{1/2}c_{1/2}} \right)
\end{equation}
and consistency with \cite{Andrade2013} can be easily verified.
Consistency between this result in $d=3$ and the massive gravity results of \cite{Blake2013} can also be seen simply by identifying $e = L = r_h = 1$, $\kappa^2 = 1/2$, and $\beta = - a_1 c_1^2$, $\alpha = - a_{1/2}c_{1/2}$. Note that the DC conductivity is not temperature independent in three dimensions, whenever the square root terms are non-vanishing; we will analyse the temperature dependence below. 

The background brane solutions coincide between our model and massive gravity. The DC conductivities agree since the fluctuation equations also coincide for homogeneous fluctuations carrying no spatial momenta. We show in appendix \ref{sec:appb} that the fluctuation equations in our model and in massive gravity are completely equivalent at zero frequency.

\subsection{Parameter Space Restrictions}

At this point in our analysis we need to place restrictions on the parameter space to obtain a physical model. Any phase of our system can be fully described by three dimensionless parameters: $\tau = T/\mu$, $\tilde{\beta} = 2a_1 c_1^2 / \mu^2$, and $\tilde{\alpha} = a_{1/2}c_{1/2}/\mu$ where we have used the chemical potential $\mu$ to fix the scaling symmetry. Given these values we may use \eqref{eq:temperature} to fix the horizon location $\mu z_0$:
\begin{equation}
	\mu z_0 = \begin{cases}
		\frac{4 \pi \tau + \tilde{\alpha} \pm \sqrt{(4 \pi \tau + \tilde{\alpha})^2 + 2dP^2}}{P^2} & P^2 \neq 0 \\
		\frac{d}{4\pi\tau + \tilde{\alpha} } & P^2 = 0
	\end{cases}
\end{equation}
where $P^2 = \tilde{\beta}+ \frac{(d-2)^2}{d-1}$.

Positivity of the norms of the two point functions of the scalar operator dual to the massless scalar field (or, equivalently, absence of ghosts) requires that $a_1 \ge 0$. Since $c_1$ and $\mu$ are real, $\tilde{\beta } \ge 0$ and hence $P^2 \ge 0$. The sign of $\tilde{\alpha}$ is more subtle, as it depends on $a_{1/2}$, $c_{1/2}$ and $\mu$. The non-linearity of the square root terms however prevents us from placing restrictions on the sign of $a_{1/2}$.  Previously we showed that $a_{1/2} c_{1/2}$ should be positive when $\mu = 0 = a_1$. This suggests $\td{\alpha}$ should be positive, for positive $\mu$, and negative for negative $\mu$.  

The $\td{\beta} \geq 0$ constraint is the only one that we can apply without direct knowledge of the sign of $\mu$. 
We now consider the cases of positive and negative $\mu$ separately, imposing the following constraints:
\begin{itemize}
	\item $T \geq 0$: The system has a non-negative temperature.
	\item $z_0 > 0$: The black brane horizon location is at a real and positive position in the holographic bulk direction.
	\item $\sigma_{DC} \geq 0$: The system has a non-negative conductivity.
	\item $f(z) > 0$ for $z \in (0,z_0)$: The point $z=z_0$ is indeed the true horizon location, no other horizons exist between this and the boundary.
\end{itemize}
We do not consider the $\mu = 0$ case here. 

For positive chemical potential, the temperature constraint $T \geq 0$ translates simply into $\tau > 0$. Imposing $z_0 > 0$ requires the root $\mu z_0^{+}$ to be the horizon location. Positive DC conductivity requires
\begin{equation}
	\frac{(d-2)^2}{\td{\beta} + \td{\alpha} (\mu z_0)^{-1}} \geq -1
\end{equation}
since we know that $z_0 > 0$ for all $\mu$ by construction. The constraint is automatically satisfied for $\td{\alpha} > 0$, i.e. $a_{1/2} c_{1/2} > 0$. The constraint can be satisfied for negative $\td{\alpha}$, but only for a finite range of temperatures. Since we wish to consider only systems which exist for arbitrary temperatures we must therefore restrict to $\td{\alpha} > 0$. 

Recall that our blackening function $F(z)$ is given by
\be
	F(z) = 1 - \frac{m_0}{\mu^d} (\mu z)^d + \frac{(d-2)(\mu z)^{2(d-1)}}{2(d-1) (\mu z_0)^{2(d-2)}} - \frac{\td{\alpha}(\mu z)}{d-1} - \frac{\td{\beta}(\mu z)^2}{2(d-2)}
\ee
where the mass parameter $m_0$ is given by
\begin{equation}
	\frac{m_0}{\mu^d} = \frac{1}{(\mu z_0)^d}\left[1 - \frac{\td{\alpha} \mu z_0}{d-1} + (\mu z_0)^2\left(- \frac{\td{\beta}}{2(d-2)} + \frac{d-2}{2(d-1)}\right)\right].
\end{equation}
We must place the constraint that $F(z) > 0$ for all $0 \leq z < z_0$ to ensure that $z_0$ is in fact the true horizon of interest. This condition is equivalent to the statement that $F(z)$ has no real roots in the open interval $z \in (0,z_0)$ which we prove in the appendix.

For negative chemical potential, $\tau < 0$ and the correct choice of the horizon location is $\mu z_{0}^-$. Positive DC conductivity requires that $\td{\alpha} < 0$, so $a_{1/2} c_{1/2} > 0$. In the table below we summarise the restrictions necessary for a realistic model:
\begin{center}
	\begin{tabular}{|c|c|}
		\hline 
		$\mu > 0$ & $ \mu < 0$ \\ \hline \hline
		\multicolumn{2}{|c|}{$a_1 > 0$} \\ \hline
		\multicolumn{2}{|c|}{$a_{1/2} c_{1/2} > 0$} \\ \hline
		\multicolumn{2}{|c|}{$\td{\beta} > 0$} \\ \hline
		$\td{\alpha} \geq 0$ & $\td{\alpha} \leq 0$ \\ \hline
		$\mu z_0 = \mu z_0^{-} > 0$ & $\mu z_0 = \mu z_0^{+} < 0$ \\ \hline
	\end{tabular}
\end{center}

Note that the restrictions discussed in this section do not ensure complete thermodynamic stability as other possible phases have not been investigated here.

\subsection{DC Conductivity Temperature Dependence}
The DC conductivity of our model in terms of the dimensionless parameters is given by
\begin{equation}
	\sigma_{DC}/\mu^{d-3} = (\mu z_0)^{-(d-3)}\left(1 + \frac{(d-2)^2}{\td{\beta} + (\mu z_0)^{-1} \td{\a}}\right)
\end{equation}
The model presented in \cite{Andrade2013} found that $\sigma_{DC}$ was independent of temperature in $d=3$ at fixed $\td{\beta}$. This can indeed  be seen from the above. Due to the presence of this additional $\td{\alpha}$ term and the accompanying factor of $(\mu z_0)^{-1}$ our model is \textit{not} independent of temperature even in $d=3$.

Using \eqref{eq:temperature} one can show that
\begin{equation}
	\frac{d (\mu z_0)}{d \tau} = - \frac{8 \pi (\mu z_0)^2(d-1)}{P^2(\mu z_0)^2 + 2d}
\end{equation}
and thus $\mu z_0$ decreases monotonically with $\tau$.  In the bulk this corresponds to the location of the horizon moving towards the boundary as we go to higher temperatures.

One can also show that 
\begin{equation}
	\frac{d}{d \tau}\left( \frac{\sigma_{DC}}{\mu^{d-3}} \right) = - \frac{d(\mu z_0)}{d \tau}\left[ (d-3)\frac{\sigma_{DC}}{\mu^{d-3}} (\mu z_0)^{-1} - \frac{(d-2)^2 (\mu z_0)^{-(d-1)} \td{\a}}{(\td{\b} + (\mu z_0)^{-1} \td{\a})^2}\right]
\end{equation}
and hence we can see that, in $d = 3$, $\sigma_{DC}$ will increase (decrease) with $\tau$ if $\td{\a}$ is negative (positive).
For $d > 3$, the DC conductivity always increases with temperature.

\begin{figure}
	\includegraphics[scale=0.4]{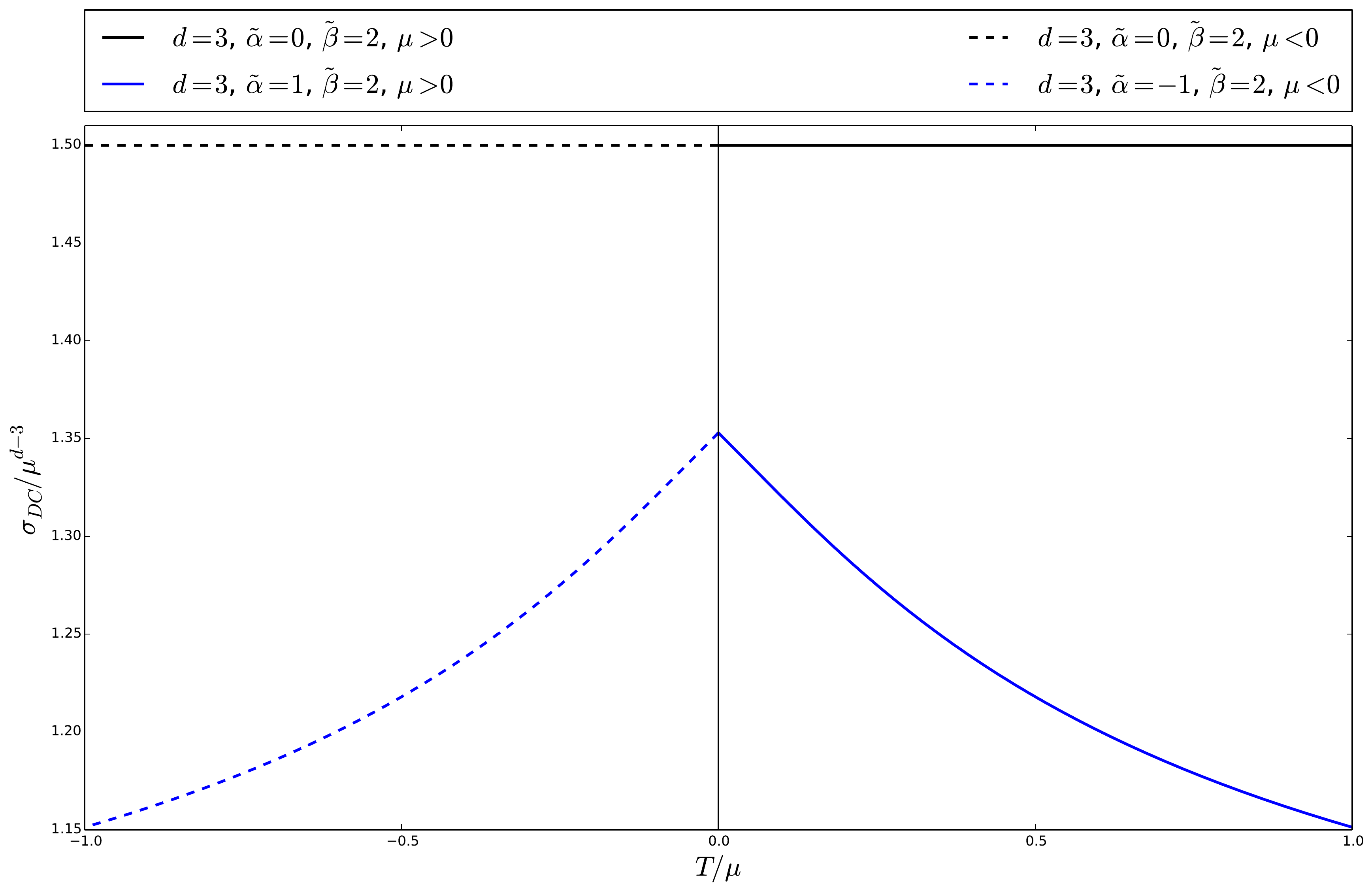}
	\centering
	\caption{Plots of $\sigma_{DC}/\mu^{d-3}$ against $T/\mu$ in $d=3$ for the given values of $\td{\alpha}$ and $\td{\beta}$. Solid lines denote results for the $\mu > 0$ branch, dashed lines denote results for the $\mu < 0$ branch. Note that $\sigma_{DC}$ decreases with $T$ for $\td{\alpha}$ non-zero.}
	\label{fig:conductivities}
\end{figure}

Figure \ref{fig:conductivities} shows a plot of $\sigma_{DC}/\mu^{d-3}$ as a function of $\tau$ for various choices of $\td{\alpha}$ and $\td{\beta}$ in $d = 3$. The DC conductivity decreases linearly with temperature for $T/\mu \lesssim 0.5$ and the slope decreases at higher temperatures. 

Recall that for the $\mu < 0$ plots decreasing $\tau$ corresponds to increasing $T$. 
The symmetry between the $\mu > 0$ and the $\mu < 0$ branches is easily understood because $\td{\a}(\mu z_0)^{-1}$ is invariant under $\mu \to - \mu$, $\td{\alpha} \to - \td{\alpha}$, $\tau \to - \tau$.

\begin{figure}
	\includegraphics[scale=0.4]{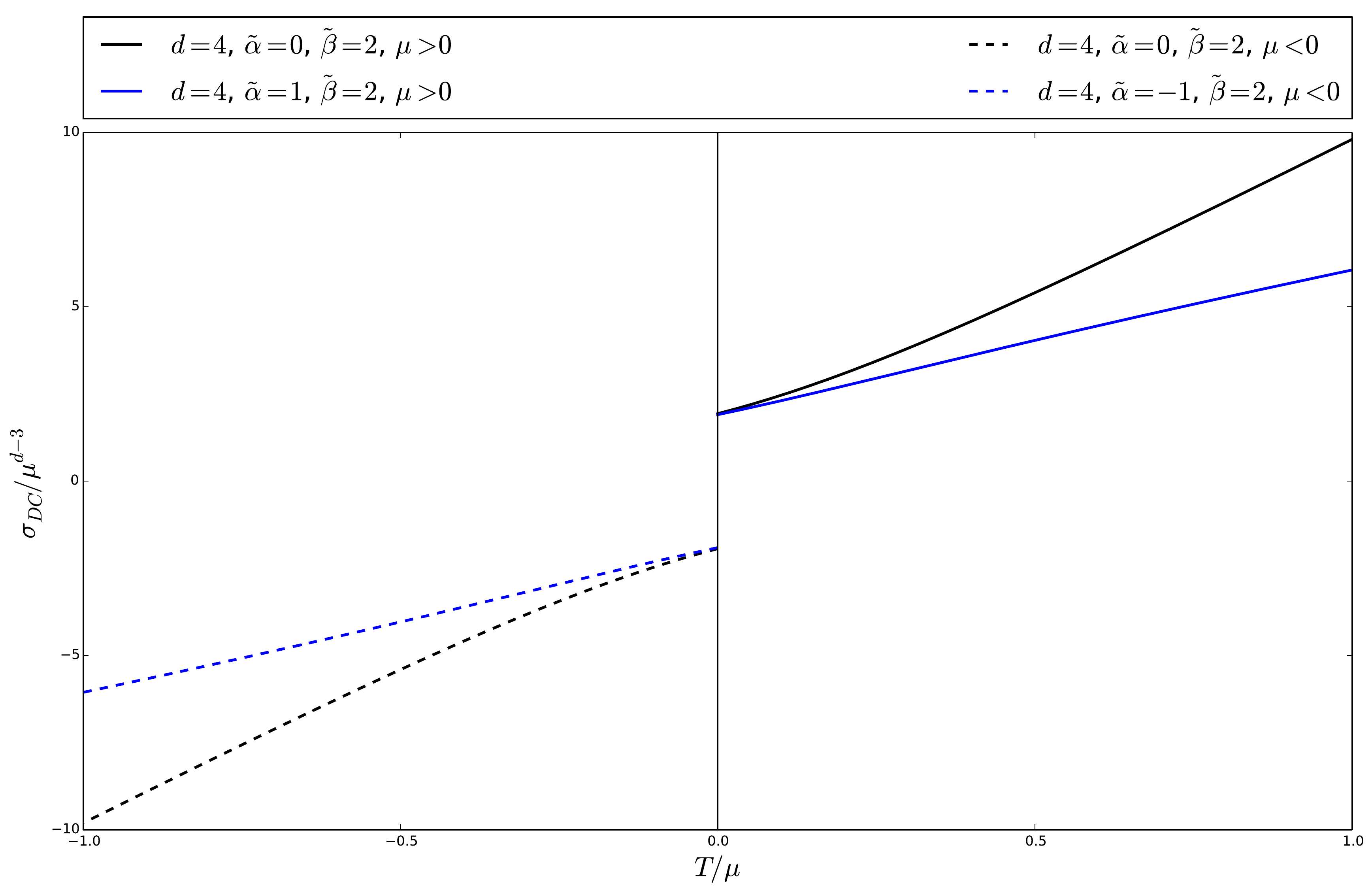}
	\centering
	\caption{Plots of $\sigma_{DC}/\mu^{d-3}$ against $T/\mu$ in $d=4$ for the given values of $\td{\alpha}$ and $\td{\beta}$. Note that $\sigma_{DC}$ is strictly increasing in $T$.}
	\label{fig:conductivities_4d}
\end{figure}

Shown in Figure \ref{fig:conductivities_4d} is a plot of $\sigma_{DC}/\mu^{d-3}$ against $\tau$ for the same choices of $\td{\alpha}$ and $\td{\beta}$ in $d=4$. Note that the reflection symmetry between the $\mu > 0$ and $\mu < 0$ branches is broken due to $\sigma_{DC}/\mu^{d-3}$ gaining a minus sign due to the odd power of $\mu$ in the $\mu < 0$ branch. We note that for $d > 3$ $\sigma_{DC}$ always increases with $T$, even in the $\td{\alpha} = 0$ case, whereas in $d = 3$ it is constant or decreases with $T$.


\subsection{Finite frequency behaviour at low temperature}

The low frequency behaviour of the AC conductivity at low temperature can be obtained by rewriting the fluctuation equations as Schr\"{o}dinger equations and matching asymptotics between IR and UV regions. This technique has been applied to a number of AdS/CMT models, see for example \cite{Horowitz:2009ij,Goldstein:2009cv,Charmousis:2010zz,Donos2012,Gouteraux:2013oca,Gouteraux2014}.

Following this framework we work in the near extremal limit and apply a matching argument to relate the IR Green's functions to the UV current-current Green's function via 
\begin{equation}
	\mathrm{Im}[G_{\mathcal{J}^x\mathcal{J}^x}^R(\omega,T)] = \sum_M d^M \mathrm{Im}[G_{\mathcal{O}_M \mathcal{O}_M}^R(\omega,T)]
\end{equation}
where $M$ runs over all the IR irrelevant operators $\mathcal{O}_M$ coupling to the current $\mathcal{J}^x$, and $d^M$ are certain numerical constants whose values are unimportant for our discussion. The operators  involved are the current itself and the two scalar operators dual to the scalar fields associated with the $x$ direction, corresponding to the perturbations $a_x, \mathcal{X}_x, \Psi_x$.

The strategy is as follows. The fluctuation equations, after decoupling, can be brought into Schr\"odinger form:
\begin{equation}
	\ddot{\bar{H}} + \omega^2 \bar{H} - V(\rho)\bar{H} = 0, \qquad V(\rho) = \frac{c_H}{\rho^2} + \ldots,
\end{equation}
where dots denote derivatives with respect to a suitably defined radial coordinate $\rho$. Expressed in this form
one can immediately extract the scaling behaviour for the imaginary part of the Green's function of the field $H_I$ with dual operator $\mathcal{O}_H$  \cite{Donos2012,Gouteraux2014}:
\begin{equation}
	\mathrm{Im}[G^R_{\mathcal{O}_H\mathcal{O}_H}(\omega\ll\mu,T=0)] \sim \omega^{\sqrt{4c_H + 1}}.
\end{equation}
The scaling behaviour of the real part of the optical conductivity is then given by
\begin{equation}
	\mathrm{Re}[\sigma(\omega\ll\mu,T=0)] = \frac{1}{\omega}\mathrm{Im}[G_{\mathcal{J}^x\mathcal{J}^x}^R(\omega\ll\mu,T=0)],
\end{equation}
where we have used the Kramers-Kronig relation, and is therefore controlled at low frequency by the lowest (IR) dimension operator. From the analysis in the previous sections we know that our system has two massless modes, i.e. two marginal operators, and we will now show that the third mode corresponds to an irrelevant operator in the IR. 

All three field equations for the linearised fluctuations involve terms of the form
\begin{equation}
	(z^{-\delta} F H')' + \frac{\omega^2}{F}H z^{-\delta}. \label{eq:non-schrodinger-derivative-term}
\end{equation}
For generic $\delta$ and $H(z)$ we can bring \eqref{eq:non-schrodinger-derivative-term} into a form more easily related to the Schr\"odinger form by making the change of coordinate $z \to \rho$ and change of variables $H(z) = z^{\delta/2}\bar{H}(\rho)$ where we define the radial coordinate as
\begin{equation}
	\frac{d \rho}{d z} = F^{-1}. \label{eq:schrodinger_rho_defn}
\end{equation}
Carrying out these substitutions yields:
\begin{equation}
	(z^{-\delta} F H')' + \frac{\omega^2}{F}z^{-\delta} H = \frac{z^{-\delta/2}}{F}\left[ \ddot{\bar{H}} + \omega\bar{H} - V_\delta(\rho)\bar{H} \right]
\end{equation}
where the potential term $V_\delta(\rho)$ is given by:
\begin{equation}
	V_\delta(\rho) = \frac{\delta}{4z^2}( (\delta + 2)F^2 - 2 z \dot{F} ).
\end{equation}
The blackening function in near-horizon (IR) limit, in the extremal case, is given by
\begin{equation}
	 F(z) = \frac{1}{2}(z-z_0)^2F''(z_0) + O((z-z_0)^3)
\end{equation}
Since the extremal limit of the black-brane solution occurs when $F(z_0) = F'(z_0) = 0$ the parameters are related as follows:
\begin{align}
	m_0 &= z_0^{-d} + \frac{\mu^2}{\gamma^2 z_0^{d-2}} - \frac{a_{1/2}c_{1/2}}{(d-1)z_0^{d-1}} - \frac{a_1 c_1^2}{(d-2)z_0^{d-2}} \\
	\frac{(d-2)\mu^2 z_0^2}{\gamma^2} &= d - a_{1/2}c_{1/2} z_0 - a_1 c_1^2 z_0^2.
\end{align}
The near horizon geometry remains $AdS_2 \times R^{d-1}$ in the presence of the scalar field profiles.

Recalling the definition \eqref{eq:schrodinger_rho_defn} of $\rho$, the Schr\"odinger coordinate, it must have the following relation to $z$ in the extremal IR limit:
\begin{equation}
	\rho = - \frac{2}{F''(z_0) (z-z_0)} + O((z-z_0)^{-2}) \label{eq:extremal_ir_rho}
\end{equation}
so the $z \to z_0$ limit corresponds to the $\rho \to \infty$ limit. In this limit, 
\begin{equation}
	F(\rho) = \frac{2}{F''(z_0)\rho^2} + O(\rho^{-3}), \qquad \dot{F}(\rho) = - \frac{4}{F''(z_0)\rho^3} + O(\rho^{-4})
\end{equation}
and thus
\begin{equation}
	V_\alpha(\rho) = \frac{2 \alpha}{F''(z_0)\rho^3 z_0} + O(\rho^4)
\end{equation}
where
\begin{equation}
	F''(z_0) = \frac{2d(d-1)}{z_0^2} - \frac{2d-3}{z_0}a_{1/2}c_{1/2} - 2(d-2)a_1 c_1^2
\end{equation}
where we have used the conditions $F(z_0) = 0$ and $F'(z_0) = 0$ to eliminate $m_0$ and $\mu^2/\gamma^2$ respectively in terms of the other parameters. Clearly $V_\alpha \sim \rho^{-3}$ in the IR limit.

After performing the change of coordinate $z \to \rho$ as discussed above, and introducing the new variables $a_I = a z^{(d-3)/2}$, $\zeta_I = \zeta z^{-d/2}$, $\xi_I = \xi z^{-(d-1)/2}$ the three field equations read, in the IR limit:
\begin{align}
	\ddot{a} + \omega^2 a &= \frac{2}{F''(z_0) \rho^2}\left[ (d-2)^2 \mu^2 a - i(d-2)\mu a_{1/2} z_0^{-1/2} \zeta - 2i(d-2)\mu a_1 c_1 \xi \right] + O(\rho^{-3}) \nn \\
	\ddot{\zeta} + \omega^2 \zeta &= \frac{2}{F''(z_0)\rho^2}\left[ i(d-2)\mu c_{1/2} z_0^{-1/2} a + a_{1/2}c_{1/2} z_0^{-1} \zeta + 2a_1 c_1 c_{1/2} \xi \right] + O(\rho^{-3})\\
	\ddot{\xi} + \omega^2 \xi &= \frac{2}{F''(z_0)\rho^2}\left[ i(d-2)\mu c_1 a + a_{1/2}c_1 z_0^{-1/2} \zeta + 2a_1 c_1^2 \xi \right] + O(\rho^{-3}). \nn
\end{align}
This system can be decoupled with the following linear combinations of fields:
\begin{align}
	\lambda_1 &= a - \frac{i a_{1/2}}{\mu z_0^{1/2}(d-2)}\zeta + i\left( \frac{(d-2)\mu}{c_1} + \frac{a_{1/2}c_{1/2}}{\mu z_0(d-2)}\right)\xi \\
	\lambda_2 &= a + i z_0^{1/2}\left(\frac{(d-2)\mu}{c_{1/2}} + \frac{2a_1 c_1^2}{\mu c_{1/2}(d-2)} \right)\zeta - \frac{2 i a_1 c_1}{\mu(d-2)}\xi \nn \\
	\lambda_3 &= a - \frac{i a_{1/2}}{\mu z_0^{1/2}(d-2)}\zeta - \frac{2i a_1 c_1}{\mu(d-2)}\xi \nn
\end{align}
which have field equations:
\begin{align}
	\ddot{\lambda}_1 + \omega^2 \lambda_1 &= O(\rho^{-3}); \\
	\ddot{\lambda}_2 + \omega^2 \lambda_2 &= O(\rho^{-3}); \nn \\
	\ddot{\lambda}_3 + \omega^2 \lambda_3 &= \frac{2}{F''(z_0)\rho^2}\left((d-2)^2 \mu^2 + 2a_1 c_1^2 + a_{1/2}c_{1/2} z_0^{-1}\right) \lambda_3 + O(\rho^{-3}). \nn
\end{align}
From these we can read off the various coefficients of interest to be:
\begin{equation}
	c_{\lambda_1} = 0, \quad c_{\lambda_2} = 0, \quad c_{\lambda_3} = \frac{2 \nu}{F''(z_0)}, \quad \nu = (d-2)^2 \mu^2 + a_{1/2}c_{1/2}z_0^{-1} + 2 a_1 c_1^2
\end{equation}
and so the IR Green's functions have the following scaling behaviour:
\begin{eqnarray}
&& 	\mathrm{Im}[G_{\lambda_1\lambda_1}^R(\omega\ll\mu, T=0)] \sim \omega, \quad	\mathrm{Im}[G_{\lambda_2\lambda_2}^R(\omega\ll\mu, T=0)] \sim \omega, \\
&&  \mathrm{Im}[G_{\lambda_3\lambda_3}^R(\omega\ll\mu, T=0)] \sim \omega^{\sqrt{8 \nu F''(z_0)^{-1} + 1}} \nn
\end{eqnarray}
Hence the dominant behaviour of the optical conductivity is
\begin{equation}
	\mathrm{Re}[\sigma(\omega \ll \mu, T = 0)] \sim
	\begin{cases}
		\omega^{\sqrt{8\nu F''(z_0)^{-1} + 1} - 1} & - F''(z_0) \leq 8 \nu < 0 \\
		1 & \nu > 0
	 \end{cases}
\end{equation}
In the previous section we derived restrictions on our parameter space and with these restrictions 
 $\nu \geq 0$ and thus the third operator (dual to $\lambda_3$) is irrelevant. Hence the dominant behaviour of the optical conductivity is controlled by the marginal operators, 
\begin{equation}
	\mathrm{Re}[\sigma(\omega \ll \mu, T = 0)] \sim 1
\end{equation}
which is consistent with metallic behaviour. In the next sections we will however show that our models do not behave as ordinary metals with sharp Drude peaks but instead display features more reminiscent of heavy fermion systems. 
 
\subsection{Relation to Drude behaviour}

As discussed in \cite{Davison:2013jba}, in Drude metals momentum is dissipated since
\be
\qquad \partial_i \langle T^{i I} \rangle  = \langle J_i \rangle F^{i I} - (\e + p) {\tau_r}^{-1} u^I.
\ee
Here $I$ denotes a spatial direction; $i$ denotes all $d$ space-time directional $J^i$ is the current; $F^{i I}$ is the gauge field strength; $\tau_r$ is the relaxation constant; $u^I$ the spatial velocity; $\e$ the energy density and $p$ the pressure. This equation reflects a loss of momentum density at a rate proportional to the velocity. Noting that in equilibrium the momentum density $P^I$ is $T^{0 I} = (\e + p) u^I$, the quantity $\tau_r$ can be interpreted as the momentum relaxation timescale; the equation above is the the covariant generalisation of 
\be
\frac{d P^I}{dt} = q E^I - \frac{P^I}{\tau_r} \label{diss}
\ee
with $q$ the charge density and $E^I$ the electric field.
In such a model the optical conductivity takes the Drude form, namely
\be
\sigma(\omega) = \frac{\sigma_{DC}}{(1 - i \omega \tau_r)}
\ee
where $\tau_r$ is the relaxation time given above and the DC conductivity is $\sigma_{DC}$.  

In the models analysed here, momentum relaxation is governed by the Ward identity
\be
\nabla^{i} \langle T_{ij} \rangle = \langle J^{i} \rangle F_{ij} + \sum_{I=1}^{d-1} \left ( \partial_j \psi_{(0)I} \langle {\cal O}_{\psi_{I}} \rangle + \partial_j \chi_{(0)I} \langle {\cal O}_{\chi_{I}} \rangle \right ). 
\ee
In the equilibrium black brane configurations the gauge field strength of the source $F_{ij}$ is zero and the expectation values of the scalar operators vanish. Working to linearised order in the perturbations 
\be
\partial^i \langle \delta T_{ij} \rangle = q \delta F_{t j} + \sum_{I=1}^{d-1} \delta_{jI} \left ( c_{1/2} \langle \delta {\cal O}_{\psi_{I}} \rangle + c_{1} \langle \delta {\cal O}_{\chi_{I}} \rangle \right ),
\ee
where $q$ is the background charge density, defined below \eqref{charge}. The time component of this identity reduces to
\be
\partial_i \langle \delta T^{i0} \rangle = 0,
\ee
so energy is conserved, but momentum is dissipated since 
\be
\partial^i \langle \delta T_{iI} \rangle = q \delta E_I + \left ( c_{1/2} \langle \delta {\cal O}_{\psi_{I}} \rangle + c_{1} \langle \delta {\cal O}_{\chi_{I}} \rangle \right ). \label{eq:ward}
\ee
The operator expectation values can be expressed in terms of terms in the asymptotic expansions near the conformal boundary as follows:
\bea
\langle \delta T_{tI} \rangle &=& d ( \delta g_{(d) t I}) = d e^{-i \omega t} H_{(d) t I}; \\
\delta E_I &=& \partial_{t} \delta A_{(0) I} = - i \omega e^{-i \omega t} a_{(0) I};  \nn \\
\langle \delta {\cal O}_{\psi_{I}} \rangle &=& a_{1/2} \frac{(d+1)}{c_{1/2}} \delta\psi_{(d+1)I} =a_{1/2} \frac{(d+1)}{c_{1/2}} e^{-i \omega t} \Psi_{(d+1) I}; \nn \\
\langle \delta {\cal O}_{\chi_{I}} \rangle &=& 2 a_1 d \delta \chi_{(d )I} = 2 a_1 d e^{-i \omega t} \mathcal{X}_{(d) I}. \nn
\eea
The expressions for the stress energy tensor and the operators dual to the square root fields follow from linearising the expressions given in \eqref{hol:d} (with $16 \pi G_{d+1} = 1$). The metric and scalar field perturbations are expressed in frequency modes in \eqref{fluc:mode}; $H_{(n) t I}$ refers to the coefficient of the $z^n$ term in the asymptotic expansion as $z \rightarrow 0$. The expressions for the expectation values of the operators dual to the massless scalar fields follow from those given in \cite{Skenderis:2002wp}, taking into account the non-canonical normalisations of the fields.

The Ward identity \eqref{eq:ward} can therefore be expressed in terms of the following algebraic relation between terms in the asymptotic expansions of the fields:
\be
i d \omega H_{(d) tI} = i \omega q a_{(0) I} + (d+1) a_{1/2} \Psi_{(d+1) I} + 2 d a_1 c_1 \mathcal{X}_{(d)I}. 
\ee
This identity is the leading order component of the equation \eqref{eq:mm_einsten} as $z \rightarrow 0$; recall that the diffeomorphism Ward identity follows from the $(zI)$ Einstein equation, which is equivalent to the $(tI)$ Einstein equation \eqref{eq:mm_einsten}. This equation is only of the form \eqref{diss} if the last two terms are proportional to the momentum density, i.e. $H_{(d) tI}$, with a real coefficient of proportionality. In the linearised limit, all normalizable modes are proportional to $a_{(0)I}$ but the constants of proportionality depend on the frequency and are complex. There is no guarantee that in the $\omega \rightarrow 0$ limit the expression above can be written in the form \eqref{diss} with a real relaxation constant. As we discuss in the next section, fitting the conductivity in our model to the Drude form requires a complex relaxation constant, i.e. momentum oscillations as well as dissipation. 

\subsection{AC conductivity numerics}

In this section we explore the behaviour of the AC conductivity by numerically solving the linearised perturbations equations.
To find the values of $\sigma(\omega)/\mu^{d-3}$ numerically we use a Mathematica code to solve the shooting problem of solving these ODEs with the desired near-boundary asymptotics and in-going boundary conditions at the horizon.
The code calculates  the $r$ series expansions of the dimensionless perturbations near the horizon and the boundary with some randomly chosen initial data. This initial data is then used in Mathematica's \texttt{NDSolve} function to integrate the ODEs to some pre-determined point in the bulk. At that point the difference between the perturbations and their first derivatives coming from the two ends is computed.
The process is then repeated for some initial data that is close to the randomly chosen data to construct an approximation to the Jacobian. We then proceed via the multivariate secant method of root finding to find initial data that is a better approximation to the true data that causes the difference function to vanish. We analysed the case of $d=3$ but qualitatively similar behaviour is likely to occur in other dimensions. 

\begin{figure}
	\centering
	\includegraphics[scale=0.3]{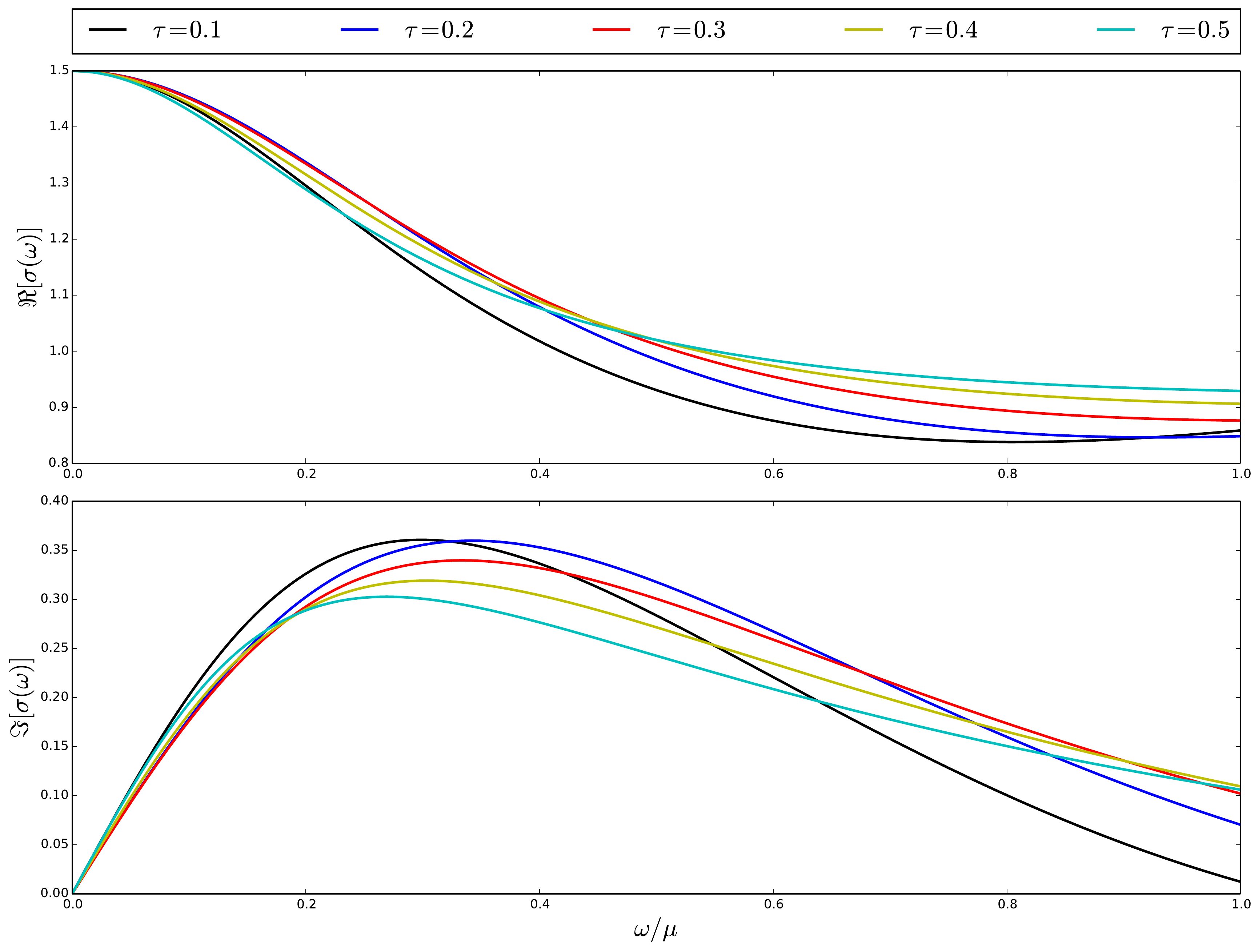}
	\caption{AC conductivity in $d = 3$ for $\td{\alpha} = 0, \td{\beta} = 2$.}
	\label{fig:ac_conductivity_alpha-2_beta0}
\end{figure}
\begin{figure}
	\centering
	\includegraphics[scale=0.3]{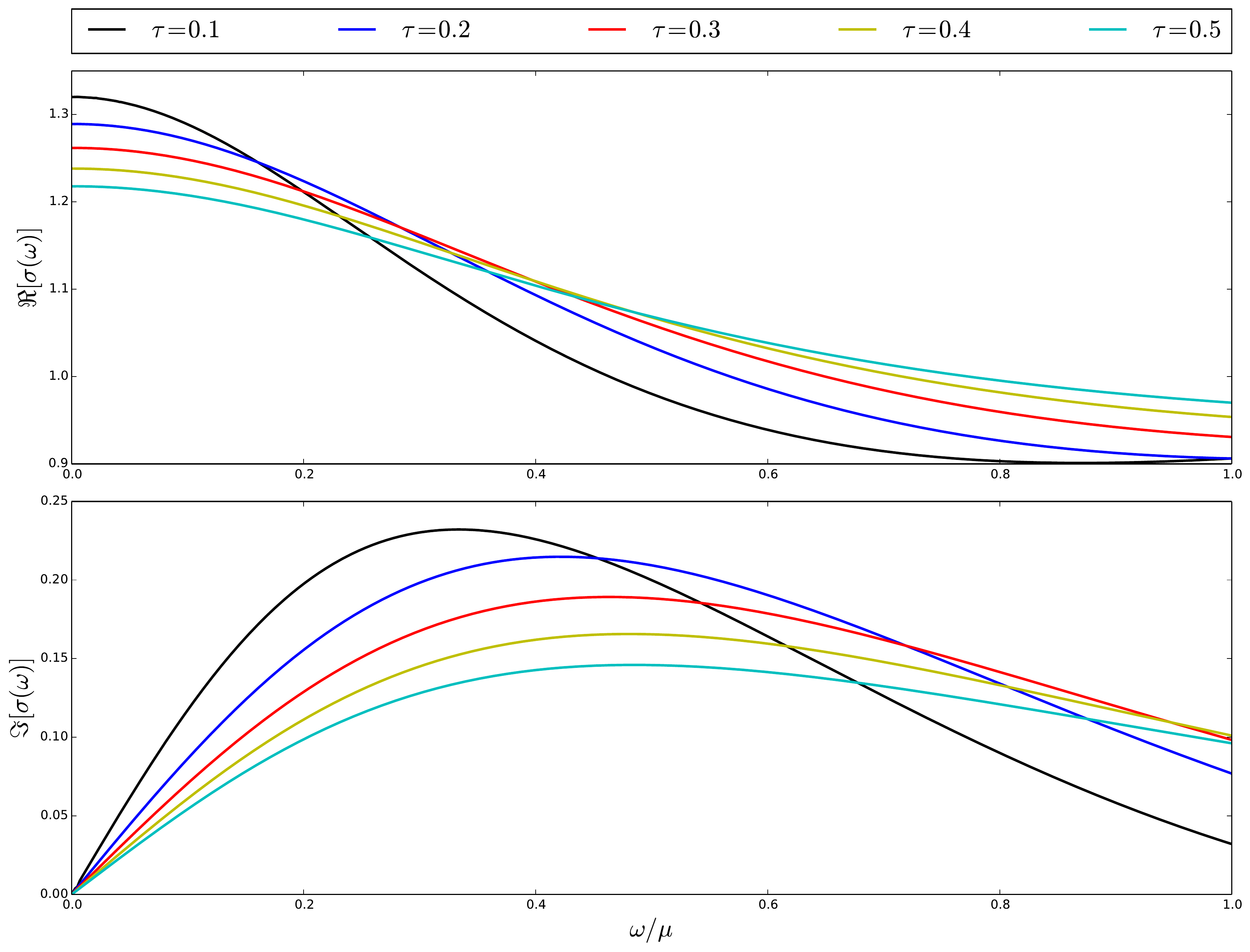}
	\caption{AC conductivity in $d = 3$ for $\td{\alpha} = 1, \td{\beta} = 2$.}
	\label{fig:ac_conductivity_alpha-2_beta-1}
\end{figure}
\begin{figure}
	\centering
	\includegraphics[scale=0.3]{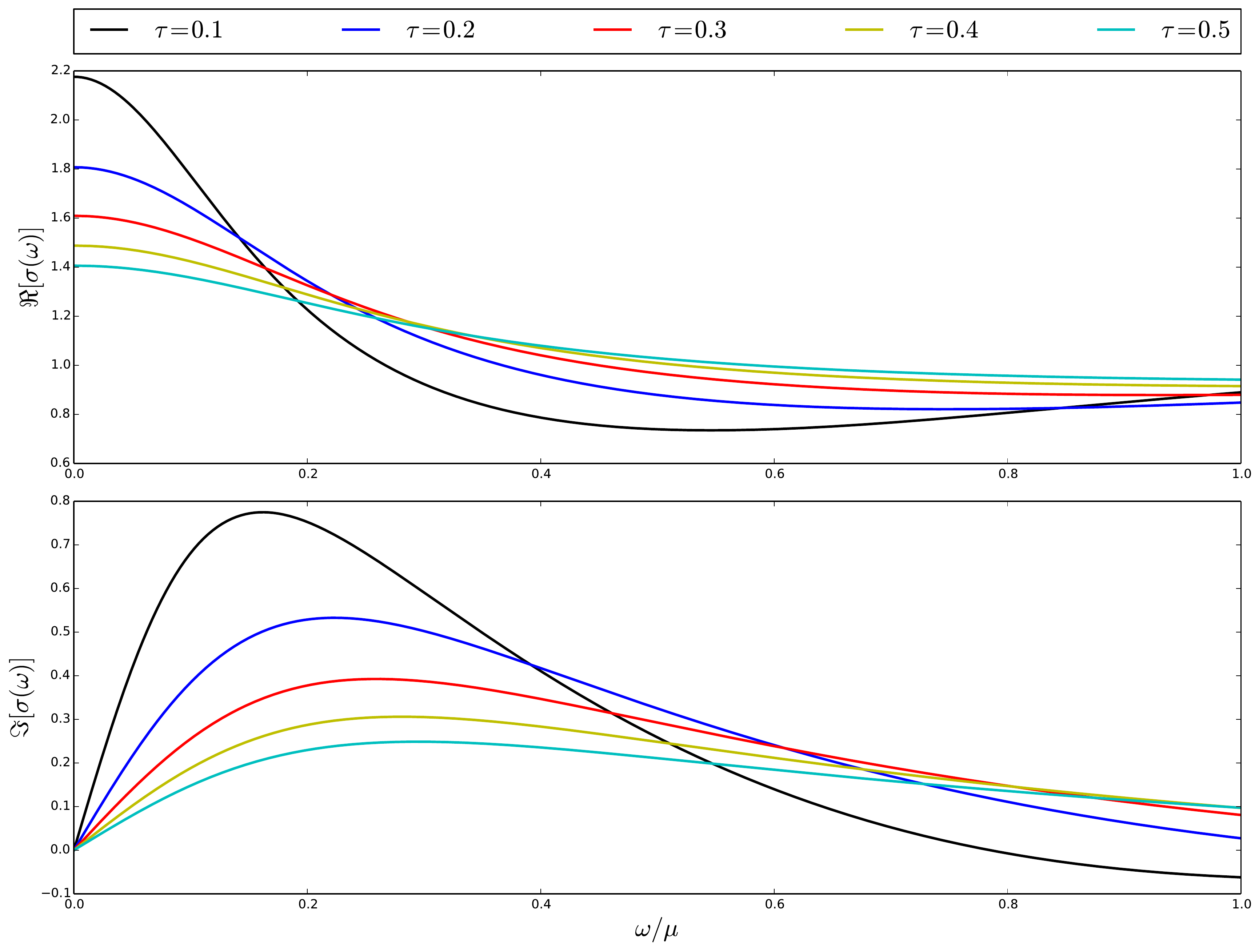}
	\caption{AC conductivity in $d = 3$ for $\td{\alpha} = 1, \td{\beta} = 0$.}
	\label{fig:ac_conductivity_alpha0_beta-1}
\end{figure}

Included in Figures \ref{fig:ac_conductivity_alpha-2_beta0}, \ref{fig:ac_conductivity_alpha-2_beta-1}, and \ref{fig:ac_conductivity_alpha0_beta-1} are plots of our numerical results for the temperatures $\tau = 0.1,\,0.2,\,0.3,\,0.4$, and $\tau = 0.5$ with various model parameters, in all cases in $d = 3$. The numerical values of $\sigma(0)$ show good agreement with our analytic expression for $\sigma_{DC}$, with no difference above the scale of accuracy set by our integration.

Using the numerical results one can also investigate the fit to a Drude peak at low frequency. The numerics show that one can only fit to a Drude formula using a relaxation time $\tau_r$ which is complex; therefore our system does not behave as a Drude metal even at very low temperature. 

Unlike \cite{Horowitz:2012ky,Horowitz:2012gs}, we see no clear signs of scaling behaviour of the optical conductivity at intermediate frequencies, $T <  \omega < \mu$. 
The AC conductivity displays several features similar to that of heavy fermion compounds. Heavy fermion materials also have a DC resistivity which increases with temperature, with a transition from normal metal behaviour to hybridised behaviour occurring below the decoherence temperature. In the hybridised phase f-electrons hybridise with conduction electrons, leading to an enhanced effective mass and a hybridisation gap. Figure  \ref{fig:ac_conductivity_alpha0_beta-1} shows that the peak in the conductivity sharpens at low temperatures, and a minimum in the conductivity develops for $\tau \lesssim 0.2$ at intermediate frequencies $\omega/\mu \sim 0.5$. The minimum is enhanced by increasing $\td{\alpha}$ and decreasing $\td{\beta}$ (i.e. increasing the amplitudes of the square root scalar fields and decreasing the amplitudes of the massless scalar fields). In our models the minima in the conductivity are strong coupling phenomena, with the reduced conductivity being associated with increased amplitudes of the scalar field fluctuations at these frequencies.

\section{Generalized phenomenological models} \label{sec:five}

In this section we consider other phenomenological models based on actions with massless scalar fields and square root terms. 

\subsection{Scalar fields identified}

As we noted earlier, 
our results for the DC conductivity replicate the massive gravity results and indeed extend them to $d \geq 3$. To compare further with massive gravity we should identify our two sets of scalar fields: $\psi_I = \chi_I$. At the level of the action this is just a simple substitution:
\begin{equation}
	S = \int_{\mathcal{M}}\dd^{d+1}\sqrt{-g}\left(R + d(d-1) - \frac{1}{4}F^2 - \sum_{I = 1}^{d-1}(a_{1/2}\sqrt{(\partial \psi_I)^2} + a_1(\partial \psi_I)^2)\right)
\end{equation}
and similarly for the Einstein equations:
\begin{align}
	R_{\mu\nu} &= -d g_{\mu\nu} + \frac{1}{2}\left(F_{\mu \lambda}{F_{\nu}}^\lambda - \frac{1}{2(d-1)}F^2g_{\mu\nu}\right) \\
	&+ \sum_{I = 1}^{d-1}\left[ \frac{a_{1/2}}{2\sqrt{(\psi_I)^2}}\left(\partial_\mu \psi_I \partial_\nu \psi_I + \frac{1}{d-1}(\psi_I)^2g_{\mu\nu}\right) + a_1 \partial_\mu \psi_I \partial_\nu \psi_I\right] \nn
\end{align}
with the Maxwell equations being unchanged. The field equations for the scalar fields become:
\begin{equation}
	\nabla_\mu\left[ \left( 2 a_1 + \frac{a_{1/2}}{\sqrt{(\partial \psi_I)^2}} \right) \nabla^\mu \psi_I \right] = 0
\end{equation}
which clearly reduce to the field equations for the independent fields case when either one of $a_{1/2}$ or $a_1$ vanishes. We shall make the same ansatz for the black brane as earlier, with the blackening function being
\begin{equation}
	F(z) = 1 - m_0 z^d + \frac{(d-2)}{2(d-1)}(\mu z_0)^2 \left(\frac{z}{z_0}\right)^{2(d-1)} - \frac{\td{\alpha} (\mu z)}{d-1} - \frac{\td{\beta} (\mu z)^2}{2(d-2)}
\end{equation}
where now $\td{\beta} = 2a_1 c^2/\mu^2$ and $\td{\alpha} = a_{1/2}c/\mu$, and $\psi_I = c x^I$. This $F(z)$ was to be expected as at the level of the background spacetime imposing $\psi_I = \chi_I$ is equivalent to 
imposing $c_1 = c_{1/2} = c$.

Now consider homogeneous finite frequency perturbations around this background as before. 
The perturbation analysis for the metric and Maxwell fields is effectively unchanged whereas the scalar field perturbations $\psi_I = c x^I + e^{-i\omega t}\Psi_I(z)$ now yield:
\begin{align}
	0 &= \left( \frac{a_{1/2}}{cz} + 2a_1 \right)\left[ \Psi_I'' + \left(\frac{F'}{F} - \frac{d-1}{z}\right)\Psi_I' + \frac{\omega^2}{F^2}\Psi_I - \frac{i\omega c}{F^2} H_{tI} - c \left\{ H_{zI}' + \left(\frac{F'}{F} - \frac{d-1}{z}\right) H_{zI} \right\} \right] \nonumber \\
	&- \frac{a_{1/2}}{cz^2}(\Psi_I' - c H_{zI})
\end{align}
which reduces to either of the previous two perturbed scalar field equations in the appropriate limits. We shall once again work in the gauge $H_{zI} = 0$. The other two perturbation equations are simply:
\begin{align}
	\frac{i \omega}{F} H_{tI}' - c\left(\frac{a_{1/2}}{cz} + 2a_1\right)\Psi_I' - \frac{i (d-2) \mu \omega z^{d-1}}{F z_0^{d-2}}a_I &= 0; \\
	a''_I + \left[ \frac{F'}{F} - \frac{d-3}{z}\right] a_I' + \frac{\omega^2}{F^2} a_I - \frac{(d-2)\mu z^{d-3}}{F z_0^{d-2}}H_{tI} &= 0. \nn
\end{align}
We can again eliminate $H_{tI}$ from the scalar perturbation equations by defining a new variable
\begin{equation}
	\bar{\xi}_I = \omega^{-1} z^{-(d-1)} \left( \frac{a_{1/2}}{cz} + 2a_1 \right) F \Psi_I'
\end{equation}
which yields the following pair of field equations:
\begin{align}
&	z^{d-3}(z^{-(d-3)} F a_I')' + \frac{\omega^2}{F} a_I = \frac{(d-2)^2 \mu^2}{z_0^{2(d-2)}}z^{2(d-2)} a_I + \frac{i(d-2)\mu c}{z_0^{d-2}}z^{2(d-2)} \bar{\xi}_I; \\
&	\left(\frac{2a_1 cz + a_{1/2}}{c z^d}\right)\left( \frac{c z^d}{2a_1 c z + a_{1/2}} F \bar{\xi}_I' \right)' + \frac{\omega^2}{F} \bar{\xi}_I =
	- \frac{2a_1 c z + a_{1/2}}{cz}
	\left[ \frac{i(d-2) \mu c}{z_0^{d-2}} a_I - c^2 \bar{\xi}_I \right], \nn
\end{align}
which reduce to the equations found earlier and in previous works \cite{Andrade2013,Gouteraux2014} in the appropriate limits.

The mass matrix has vanishing determinant and as such one massless mode can be found. Consider the following combination of fields:
\begin{align}
	\lambda_{1I} &= \frac{1}{B(z)}\left[ a_I - \frac{i(d-2)\mu z^{2d-3}}{z_0^{d-2}(2a_1 cz + a_{1/2})} \bar{\xi}_I \right]; \\
	\lambda_{2I} &= \frac{1}{B(z)}\left[ \frac{(d-2)^2 \mu^2 z^{2d-3}}{z_0^{2(d-2)} (2a_1 c^2 z + a_{1/2}c)}a_I + \frac{i (d-2) \mu z^{2d-3}}{z_0^{d-2}(2a_1 c z + a_{1/2})} \bar{\xi}_I \right],  \nn
\end{align}
where the coefficient function $B(z)$ is given by
\begin{equation}
	B(z) = 1 + \frac{(d-2)^2 \mu^2 z^{2d-3}}{z_0^{2(d-2)}(2a_1 c^2 z + a_{1/2}c)}.
\end{equation}
The field equations for $\lambda_{1I}$ read
\begin{equation}
\left( z^{-(d-3)} F a_I' - \frac{i(d-2)\mu z^d}{(2a_1 cz + a_{1/2}) z_0^{d-2}} F \bar{\xi}_I' \right)' + \frac{z^{3-d} \omega^2}{F}\left(a_I - \frac{i(d-2)\mu z^{2d-3}}{z_0^{d-2}(2a_1 cz + a_{1/2})} \bar{\xi}_I \right) = 0
\end{equation}
or, equivalently,
\begin{equation}
z^{d-3} \Pi_I' + \frac{\omega^2 B}{F} \lambda_{1I} = 0
\end{equation}
where $\Pi_I$ is given by
\begin{equation}
	\Pi_I = z^{-(d-3)}F a_I' - \frac{i(d-2)\mu z^d}{(2a_1 c z + a_{1/2}) z_0^{d-2}} F \bar{\xi}_I'
\end{equation}
and is radially conserved in the limit $\omega \to 0$.

For  convenience we can rewrite $\Pi_I$ in terms of the modes $\lambda_{1I}$ and $\lambda_{2I}$ using the following relations:
\begin{align}
	a_I &= \lambda_{1I} + \lambda_{2I}; \\
	\bar{\xi}_I &= \frac{i(d-2)\mu}{z_0^{d-2} c} \lambda_1 - \frac{i z_0^{d-2}(2a_1 c z + a_{1/2})}{\mu(d-2) z^{2d-3}}. \nn \lambda_2
\end{align}
With these relations one can show that $\Pi$ is given by:
\begin{equation}
	\Pi_I = z^{-(d-3)} F B \lambda_{1I} + (2d-3) z^{-(d-2)} \lambda_{2I} + \frac{2a_1 c z}{2a_1 c z + a_{1/2}} z^{-(d-2)} \lambda_{2I}.
\end{equation}
Notice that, in the case $a_{1/2} = 0$, this reduces to the previously known conserved quantity of \cite{Andrade2013}.

The asymptotic and near-horizon analysis performed earlier is largely unchanged. We already know the asymptotic behaviour of the Maxwell perturbation $a_I$:
\begin{equation}
	a_I = a_I^{(0)} + \frac{\langle J_I \rangle e^{i \omega t}}{d-2} z^{d-2} + \ldots
\end{equation}
from our earlier analysis. It is also clear that, as long as $a_{1/2} \neq 0$, $\bar{\xi}_I$ has one normalizable mode $z^0$, and one non-normalizable mode $z^{-(d-1)}$. In the case $a_{1/2} = 0$ this non-normalizable mode becomes $z^{-(d-2)}$ and the analysis reduces to that in the previous work of \cite{Andrade2013}. Again we wish to turn off boundary sources for these perturbations, so we turn off non-normalizable modes. The asymptotic behaviour is thus given by
\begin{equation}
	\bar{\xi}_I = \bar{\xi}_I^{(0)} + O(z)
\end{equation}
The coefficient $1/B(z)$ has asymptotic behaviour given by
\begin{equation}
	\frac{1}{B(z)} = 1 - \frac{(d-2)^2 \mu^2}{c a_{1/2} z_0^{2(d-2)}} z^{2d-3} + \ldots 
\end{equation}
and hence the massless mode $\lambda_{1I}$, and conserved quantity $\Pi_I$ have the following asymptotic forms
\begin{equation}
	\lambda_{1I} = a_I^{(0)} + \ldots \qquad 
	\Pi_I = \langle J_I \rangle e^{i \omega t} + \ldots
\end{equation}
The near-horizon behaviour of the Maxwell field is unchanged:
\begin{equation}
	a_I = (z - z_0)^{i \omega / F'(z_0)}[a_I^H + O((z-z_0))]
\end{equation}
Making a similar ansatz as earlier one can deduce that, to leading order, we have
\begin{equation}
	\bar{\xi}_I = (z-z_0)^{i \omega / F'(z_0)}[ \bar{\xi}_I^H + O((z-z_0))]
\end{equation}
Recalling that the optical conductivity is defined by \eqref{eq:optical_conductivity} and that the DC conductivity is the $\omega \to 0$ limit of this we define the following auxiliary quantity:
\begin{equation}
	\sigma_{DC}(z) = \lim_{\omega \to 0} \frac{\Pi_I}{i \omega \lambda_{1I}}.
\end{equation}
Following the same steps as earlier we can show that this quantity is radially conserved and thus we may evaluate it on the horizon to find its value. Clearly $\sigma_{DC} = \lim_{z \to 0} \sigma_{DC}(z)$ from the above analysis. Hence the DC conductivity for this model is given by
\begin{equation}
	\sigma_{DC} = \sigma_{DC}(z_0) = z_0^{-(d-3)}B(z_0) = z_0^{-(d-3)}\left( 1 + \frac{(d-2)^2 \mu^2}{2a_1 c^2 + z_0^{-1}a_{1/2} c} \right)
\end{equation}
This is consistent with our earlier result where the two sets of scalars were treated as independent fields and it is also consistent with \cite{Andrade2013,Blake2013}.

To understand the behaviour of the optical conductivity we first consider the low temperature, low frequency behaviour.
In the extremal limit one can express the fluctuation equations near the horizon as
\begin{align}
	\ddot{a}_I + \omega^2 a_I &= \frac{2}{F''(z_0) \rho^2}\left[ (d-2)^2 \mu^2 a_I - i(d-2)\mu c z_{0}^{d-2} \bar{\xi}_I 
	\right] + O(\rho^{-3}) \\
	\ddot{\bar{\xi}}_I + \omega^2 \bar{\xi}_I &= - \frac{2 (2 a_1 + \frac{a_1/2}{cz_0}) }{F''(z_0)\rho^2}\left[ i(d-2)\mu c z_0^{2-d} a_I - c^2 \bar{\xi}_I \right] + O(\rho^{-3}) \nn
	\end{align}
These equations are diagonalised by the combinations 
\be
\lambda_{1I} = a_{I} - \frac{(d-2) \mu z_0^{d-2}}{ (2a_1 c + a_{1/2} z_0^{-1})} \bar{\xi}_I; \qquad 
\lambda_{2I} = a_{I} + \frac{i z_0^{d-2}}{\mu (d-2)} \bar{\xi}_I,
\ee	
resulting in 
\bea
\ddot{\lambda}_{1I} + \omega^2 \lambda_{1I} &=& O(\rho^{-3}) \\
	\ddot{\lambda}_{2I} + \omega^2 \lambda_{2I} 
	&=& \frac{2}{F''(z_0)\rho^2}\left((d-2)^2 \mu^2 + 2a_1 c^2 + a_{1/2} c  z_0^{-1}\right) \lambda_{2I} + O(\rho^{-3}). \nn
\eea
Therefore one obtains one massless mode and one (IR) irrelevant mode, whose dimension is as before, with the identification $c_1 = c_{1/2} = c$. The massless mode controls the conductivity, which therefore has a peak at zero frequency. Since the fluctuation equations are similar to those in the previous section, we would expect qualitatively similar behaviour in this model. 

\subsection{Other square root models}

The final model we will consider is 
\begin{equation}
	S = \int_\mathcal{M} \dd^{d+1} x \sqrt{-g}\left(R + d(d-1) - \frac{1}{4}F^2 - a_1\sum_{I=1}^{d-1} (\partial \chi_I)^2 - a_{1/2} \sqrt{\sum_{I=1}^{d-1}(\partial\psi_I)^2}
	\right).
\end{equation}
In general the scalar fields can no longer be considered independently of each other, unlike the previous case. The blackening function in this model is
\begin{equation}
	F(z) = 1 - m_0 z^d + \frac{(d-2)^2\mu^2}{2(d-1)} \frac{z^{2(d-1)}}{z_0^{2(d-2)}} - \frac{a_1 c_1^2 z^2}{d-2} - \frac{a_{1/2}c_{1/2}z}{(d-1)^{3/2}}
\end{equation}
which is consistent with that of massive gravity in $d = 3$.
Now let us consider perturbing the background solutions:
\begin{equation}
	\psi_I \to \psi_I + \delta \psi_I,
\end{equation}
with corresponding perturbations of the gauge field and metric. 
At the level of perturbation analysis, and of the background metric, the change $\sum_I \sqrt{(\partial \psi_I)^2} \to \sqrt{\sum_I (\partial \psi_I)^2}$ is equivalent to the rescaling $a_{1/2} \to a_{1/2}/(d-1)^{1/2}$. We can show this as follows. The two Lagrangians are
\begin{equation}
	\mathcal{L}_1 = a_{1/2}\sum_{I=1}^{d-1}\sqrt{(\partial \psi_I)^2}, \qquad \mathcal{L}_2 = a_{1/2}' \sqrt{\sum_{I=1}^{d-1}(\partial\psi_I)^2}.
\end{equation}
When one evaluates these Lagrangians onshell with the values $\psi_I = c x^I + \delta \psi_I(t,z)$ to leading quadratic order in the perturbations $\delta\psi_I$ one finds:
\begin{equation}
	\mathcal{L}_1 = a_{1/2}\left( (d-1)cz + \frac{1}{2cz}\sum_{I=1}^{d-1}(\partial\delta\psi_I)^2 \right), \qquad \mathcal{L}_2 = \frac{a_{1/2}'}{(d-1)^{1/2}} \left( (d-1) cz + \frac{1}{2cz}\sum_{I=1}^{d-1}(\partial\delta\psi_I)^2 \right)
\end{equation}
which are clearly equivalent under the identification $a_{1/2} = a_{1/2}'/(d-1)^{1/2}$.
Any result we found earlier for the model of section \ref{sec:five} can therefore be applied to this model with a rescaling of $a_{1/2}$.

\section{Conclusions} \label{sec:six}

In this paper we have focussed on simple models of explicit translational symmetry breaking. The main advantage of these models is that the brane backgrounds are isotropic and homogenous, and can therefore be constructed analytically. The holographic duals to the bulk symmetry breaking can also be explicitly identified, unlike in massive gravity models, and correspond to switching on spatial profiles for marginal couplings in the field theory. 

Couplings growing linearly with spatial directions represent a qualitatively different mechanism for momentum dissipation than lattice and phonon effects in an ordinary metal. It is therefore perhaps unsurprising that our models do not exhibit ordinary metal behaviour. Nonetheless these models do show a peak in the optical conductivity at zero frequency; the DC resistivity increases linearly in temperature at low temperature in three boundary dimensions and by tuning the parameters one obtain minima in the optical conductivity at finite frequency. These features are reminiscent of strange metals and heavy fermion systems and suggest that it may be interesting to explore such models further. 

The novel phenomenology is associated with the square root actions \eqref{eq:sqr_act}: when this term is switched off one does not find linear growth of the DC resistivity with temperature, for example. Despite the apparent non-locality of this action, we showed in section \ref{sec:three} that the holographic dictionary is well-defined and one can work perturbatively about any background solution for this action. Moreover, we can view \eqref{eq:sqr_act} as a scaling limit of a brane action \eqref{eq:brane_act}. Brane actions exhibit no non-analytic behaviours when the background field profiles vanish and should give qualitatively similar phenomenological behaviour to \eqref{eq:sqr_act}. It would therefore be interesting to develop top-down phenomenological models based on branes, which capture the desirable features of \eqref{eq:sqr_act}. 

One issue with our black brane backgrounds is that they have finite entropy at zero temperature, indicating that they may not be the preferred phase at very low temperatures. Generic Einstein-Maxwell-dilaton models admit Lifshitz and hyperscaling violating solutions whose entropy scales to zero at zero temperature, see \cite{Kachru:2008yh,Taylor:2008tg,Goldstein:2009cv,Charmousis:2010zz,Huijse:2011ef,Dong:2012se,Harrison:2012vy,Bhattacharya:2012zu,Kundu:2012jn,Iizuka:2012pn,Gath:2012pg,Lucas:2014zea}, and it would be straightforward to extend our discussion of translational symmetry breaking using massless and square root scalar fields to such models.

\section*{Acknowledgements}

Both M.T. and W.W. acknowledge support from STFC. 
M.T. acknowledges support from a grant of the John Templeton Foundation. The opinions in this publication are those of the authors and do not necessarily reflect the views of the John Templeton Foundation. 

\newpage


\appendix

\section{Blackening function roots} \label{sec:appa}

\begin{theorem}
	Let $F(z)$ be the smooth polynomial obeying the Einstein equation and the constraints $F(0) = 1$, $F(z_0) = 0$, $F'(z_0) = -4\pi T \leq 0$, 
	with $\td{\beta} > 0$ and $\td{\alpha}/(\mu z_0) > 0$. If $z_c$ is a root of $F(z)$ in the open interval $(0, z_0)$ then $F'(z_c) < 0$.
\end{theorem}
\begin{proof}
Using the Einstein equation one can prove that at any root $z_c$ of $F(z)$:
	\begin{equation}
		z_c F'(z_c) = -d + \frac{(d-2)}{2(d-1)}(\mu z_0)^2 \left(\frac{z_c}{z_0}\right)^{2(d-1)} + \frac{\td{\alpha} \mu z_c}{d-1} + \frac{\td{\beta} (\mu z_c)^2}{2(d-2)}
	\end{equation}
	and hence one can show that
	\begin{eqnarray}
		z_c F'(z_c) - z_0 F'(z_0) &=& \frac{(d-2)}{2(d-1)}(\mu z_0)^2 \left[ \left(\frac{z_c}{z_0}\right)^{2(d-1)}-1\right] + \frac{\td{\alpha} \mu z_0}{d-1}\left[ \frac{z_c}{z_0} - 1 \right]  \\
		&& \qquad + \frac{\td{\beta}(\mu z_0)^2}{2(d-2)} \left[ \left(\frac{z_c}{z_0}\right)^2 - 1\right]. \nn
	\end{eqnarray}
	Consider a root $F(z_c) = 0$ where $z_c \in (0, z_0)$. Clearly $(z_c / z_0)^n < 1$ for any positive integer $n \geq 1$. We know by assumption that $\td{\beta} > 0$, $(\mu z_0)^2$, $\td{
	\alpha} \mu z_0 > 0$, hence $z_c F'(z_c) < z_0 F'(z_0) = - 4\pi T \leq 0$. Therefore $F'(z_c) < 0$. 
	\end{proof}
\begin{lemma}
	If $T > 0$ or $F''(z_0) > 0, T = 0$ then $F(z)$ has no roots in the open interval $(0, z_0)$. (Note that $F''(z_0)$ is automatically positive at $T=0$ given the constraints of the previous theorem.)
\end{lemma}
\begin{proof}[Proof of lemma]
We know that, since $F'(z_0) \leq 0$ and $F''(z_0) > 0$, $F(z)$ must be positive before the root. Since we also know that $F(0) = 1$ is positive, there must be an even number of odd multiplicity roots and there can be any number of even multiplicity roots in the open interval $(0,z_0)$ to ensure that we can continue from a positive value at $0$ to a positive value just before $z_0$.

We know that if $z_c$ is a root, then $F'(z_c) < 0$. Thus there can be no even multiplicity roots or odd multiplicity roots of multiplicity larger than 1. This is because of the fact that if $p(x) = 0$ is a polynomial with a root at $x = x_c$ of multiplicity $n$, then $p'(x)$ has a root at $x = x_c$ of multiplicity $n - 1$.

We also know that if $z_c$ is a simple root of $F(z)$ then it must have $F'(z_c) < 0$. There is no way to reconcile having a non-zero number of such roots with the requirement that $F(z)$ must be positive just before the root $z = z_0$.
\end{proof}

\section{DC conductivity and massive gravity} \label{sec:appb}

One can also obtain the DC conductivity by switching on zero frequency perturbations, i.e. working strictly in the $\omega = 0$ limit. In this case the Maxwell equation and the $(tI)$ components of the Einstein equation immediately decouple. The Maxwell equation is 
\be
a_I '' + \left[\frac{F'}{F} - \frac{d-3}{z}\right]a_I' - \frac{\mu(d-2)z^{d-3}}{F} H_{tI}' = 0
\ee
which can be rewritten as 
\be
( F z^{3-d} a_I' - \mu (d-2) H_{tI})' = 0.
\ee
The Einstein equation is 
\bea
&&- \frac{1}{2} F H_{tI}'' + \frac{F}{2z}(d-1)H_{tI}' + \left(\frac{F'}{z} - \frac{dF}{z^2}\right) H_{tI} = \frac{1}{2}(d-2)\mu z^{d-1}\left(- F a_I' + \frac{d-2}{d-1}\mu z^{d-1}H_{tI}\right) \nn \\ && \hspace{9cm} + \frac{1}{2} a_{1/2} c_{1/2}z H_{tI}. 
\eea
In the Einstein equation, the term in the second line is the contribution from the scalar field parts of the action. 

Now let us compare these equations to those arising in massive gravity. The Maxwell equation is identical and the first line in the Einstein $(tI)$ equation is the same. The contribution in the second line is replaced by the contribution from \eqref{Se1}. Linearizing the latter around the background solution gives 
\be
\delta  \bar{T}_{tI} = \frac{1}{2} m^2 \alpha_1 z H_{tI}, 
\ee
which using \eqref{eq:mg} implies that these perturbation equations match between massive gravity and the scalar model and therefore the DC conductivities must match. 

\providecommand{\href}[2]{#2}\begingroup\raggedright\endgroup

\end{document}